\documentclass[a4paper,12pt]{article}
\usepackage{float}
\usepackage{tabularx}
\usepackage{amsmath}
\usepackage{amsfonts}
\usepackage{adjustbox}
\usepackage{multirow}
\usepackage{color}
\usepackage{longtable}
\usepackage{pdflscape}
\usepackage{rotating}
\usepackage{hyperref}
\usepackage[margin=1in,footskip=0.25in]{geometry}
\title{~~~Deuteron Structure and Form Factors: Using Inverse Potentials for S-waves}
\author{Anil Khachi$^{1}$, Lalit Kumar$^{1}$, M.R. Ganesh Kumar$^{2}$, and O.S.K.S Sastri$^{1*}$\\\\
$^{1}$Department of Physics and Astronomical Sciences,\\ Central University of Himachal Pradesh Dharamshala, 176215, \\Himachal Pradesh, Bharat (India)\\\\
$^{2}$Applied Materials India Pvt Ltd., Bengaluru ,Bharat(India)\\}
\begin{document}
\maketitle
\abstract{In this paper, we determine deuteron's static properties, low energy scattering parameters, total cross-section and form factors from inverse S-wave potentials constructed using Morse function. The scattering phase shifts (SPS) at different lab energies are determined using phase function method. The model parameters are optimised using both machine learning algorithm and traditional data analysis by choosing mean squared error as cost function. The mean absolute error between experimental and obtained SPS for states $^3S_1$ and $^1S_0$ are found to be 0.35 and 0.70 respectively. The low energy scattering parameters are matching well with expected values. The contribution due to S-waves SPS towards total cross-section at various energies have been obtained and are matching well with experimental values. The analytical ground state deuteron wavefunction (DWF) is obtained by utilizing the experimental value for Quadrupole moment. Other static properties and form factors determined from obtained DWF are found to be in close agreement with experimental ones.}\\\\\
{\textbf{keywords:} neutron-proton interaction, Deuteron, inverse potential, phase shift, Phase function method, Scattering, Morse potential, Form factors}

\newpage
\section{Introduction}
\label{sec:sample1}

Study of deuteron to understand its experimentally observed static properties has been reviewed by Zhaba \cite{1} and Garcon \cite{2}. The best results are from nucleon-nucleon (NN) interaction precision potentials \cite{3,4,5,6,7,8}. All these describe the NN interaction as consisting of long range one pion exchange. The intermediate and short range interaction are either modeled using simple functional forms \cite{7} or using meson exchanges\cite{8}. At very small inter-nucleon distances, a strong repulsive core is expected due to strong anti-corelation between the nucleons. This is modeled phenomenologically using exponential functions. Finally, the model parameters are obtained by directly fitting the experimental scattering phase shifts (SPS) for various $\ell$-channels.\\
Alternatively, there has been similar success using J-matrix inversion potentials \cite{9} and N3LO \cite{10,11}. Other potentials that have been tried are Yukawa \cite{12}, Hulthen \cite{13}, Malfiet-Tjohn(MT) \cite{14}, Manning-Rosen \cite{15} renormalised one pion exchange (OPE) and two pion exchange(TPE) \cite{16}, delta \cite{17} and different variations of Woods-Saxon potential \cite{18}.  Recently, an effective potential \cite{19} for deuteron has been obtained by employing supersymmetric (SUSY) quantum mechanics approach by considering D-state wavefunction to be proportional to that of S-state.\\
We have utilised Morse function as the reference potential \cite{20,21} to guide the construction of inverse potentials \cite{22,23}. This was achieved by solving for SPS using phase function method (PFM) in an iterative manner within an optimisation algorithm.  Having obtained the inverse potentials for S-states of NN interactions, that match the experimental SPS to a very good accuracy, the question remains as to what could be deduced from them. One obvious thing, to do, was to calculate the partial and total cross-sections. A second effort would be to obtain the low energy scattering parameters. The third aspect is to consider the fact that the resultant inverse potentials are of Morse form, for which time independent Schr$\ddot{\text{o}}$dinger equation can be solved analytically for $\ell = 0$ case. Our approach in this paper is to utilise the model parameters of $^3S_1$ inverse potential to determine it's analytical wave-function and then use a simple approximation mentioned above to determine D-state wavefunction such that overall Deuteron wavefunction (DWF) is normalised while simultaneously giving rise to correct quadrupole moment as in \cite{24}. Once, the DWF is determined, the static properties and form factors for deuteron can be determined. Detailed discussion about form factors (FFs) and related experimental data can be found in paper by Ingo Sick \cite{25}     \\
Previously, while determining the inverse potential for S-state \cite{22,23}, we have utilised all available experimental data to optimise the three model parameters of the Morse function. This amounts to building the model from data as in physics based machine learning, wherein the number of data points being used for optimisation is much larger than the number of model parameters. In the traditional approach of modeling in physics, one considers only as many experimental points as the number of model parameters and then the rest of data points are predicted. Here, we introduce a comprehensive data analysis using the later approach, where in all the possible combinations (Appendix section) of experimental data points are considered and analysed, to obtain best model parameters along with uncertainties.\\ 
Hence, the main objective, in this paper, is to determine static and low energy properties, scattering cross-section as well as form factors for deuteron using machine learning algorithm (MLA) and traditional data analysis (TDA).   

\section{Methodology:}
Selg \cite{20,21} has discussed in detail the reference potential approach to obtaining inverse potentials using Morse function, given by
\begin{equation}
V_M(r) = V_{0}\left( e^{-2(r-r_m)/a_m} - 2e^{-(r-r_m)/a_m} \right)
\label{eq1}
\end{equation} 
where, model parameters $V_0$ (MeV), $r_m$ (fm) and $a_m$ (fm) reflect depth of potential, equilibrium distance at which maximum attraction is felt and shape of potential respectively.\\ 
One can use a combination of Morse potentials if needed \cite{21}. The number of bound states available must be greater than or equal to number of model parameters to be determined. Then, one can fix the three parameters of Morse function, by considering any three of them. But, deuteron has only one and hence it is not possible to fix these exactly. This is what makes the study of deuteron an extremely interesting one. \\
Morse potential has certain interesting characteristics which separates it from other phenomenological potentials. These are:
\begin{enumerate}
\item The time independent Schr$\ddot{\text{o}}$dinger for it, is solvable analytically for E $<$ 0 bound states \cite{26}.
\item Unlike other phenomenological potentials used for studying NN interactions like Hulthen \cite{13}, Malfliet-Tjon \cite{14}, Manning-Rosen \cite{15} and others for E $>$ 0, the exact analytical expression \cite{26} for scattering state phase shifts is known for $\ell = 0$ states.
\item Relatively simpler wavefunction \cite{26}, and 
\item It is a shape invariant potential \cite{27}. 
\end{enumerate}
These advantages can be utilised to analyse the $^3S_1$ bound state and $^1S_0$ scattering state of deuteron.
\subsection{Triplet S-wave bound state energy:} 
The radial time independent Schr$\ddot{\text{o}}$dinger equation (TISE), for $\ell$ = 0 (S-wave), is given by
\begin{equation}
-\frac{\hbar^2}{2\mu} \frac{d^2u(r)}{dr^2} + V_M(r)u(r) = Eu(r) 
\label{TISE}  
\end{equation}
where $\mu$ is reduced mass of neutron and proton. The analytical solution of TISE is derived by Morse \cite{28} and ground state energy expression is given by
\begin{equation}
E_0 = -\frac{\hbar^2}{2\mu a_m^2}(\lambda-1/2)^2
\label{eigenvals}
\end{equation}
where
\begin{equation}
\lambda = \sqrt{\frac{ 2 \mu V_0 a_m^2}{\hbar^2}}
\end{equation}
is called well-depth parameter and is dependent only on $V_0$ and $a_m$.

Utilizing experimental binding energy (BE) for Deuteron as, E$_0$= $-2.224589(22)$ MeV \cite{29}, $V_0$ can be expressed in terms of $a_m$ as
\begin{equation}
V_0 = \frac{\hbar^2}{2\mu a_m^2} \left(0.5 + \sqrt{\frac{2\mu(2.224589)a_m^2}{\hbar^2}}\right)^2
\label{V0constraint}
\end{equation}
To fix the other two parameters $a_m$ and $r_m$, we utilise experimental SPS. Out of an infinite set of values for $V_0$ and $a_m$ that could give rise to experimental BE, only one set in consonance with a particular $r_m$ should give rise to observed experimental SPS. To determine SPS, Morse \cite{28} suggested phase function method.
\subsection{Phase function method (\textit{PFM}):}
The Schr$\ddot{\text{o}}$dinger wave equation for a spinless particle with energy E and orbital angular momentum $\ell$ undergoing scattering is given by

\begin{equation}
\frac{\hbar^2}{2\mu} \bigg[\frac{d^2}{dr^2}+\bigg(k^2-\frac{\ell(\ell+1)}{r^2}\bigg)\bigg]u_{\ell}(k,r)=V(r)u_{\ell}(k,r)
\label{Scheq}
\end{equation}
The second order differential equation  Eq. \ref{Scheq} has been transformed to the first order non-homogeneous differential equation of Riccati type \cite{30,31} given by  

\begin{equation}
\delta_{\ell}'(k,r)=-\frac{V(r)}{k}\bigg[\cos(\delta_\ell(k,r))\hat{j}_{\ell}(kr)-\sin(\delta_\ell(k,r))\hat{\eta}_{\ell}(kr)\bigg]^2
\label{PFMeqn1}
\end{equation}
Prime denotes differentiation of phase shift with respect to distance and the Riccati Hankel function of first kind is related to $\hat{j_{\ell}}(kr)$ and $\hat{\eta_{\ell}}(kr)$ by $\hat{h}_{\ell}(r)=-\hat{\eta}_{\ell}(r)+\textit{i}~ \hat{j}_{\ell}(r)$ . In integral form the above equation can be written as

\begin{equation}
\delta(k,r)=-\frac{1}{k}\int_{0}^{r}{V(r)}\bigg[\cos(\delta_{\ell}(k,r))\hat{j_{\ell}}(kr)-\sin(\delta_{\ell}(k,r))\hat{\eta_{\ell}}(kr)\bigg]^2 dr
\end{equation}
for $\ell = 0$,the Riccati-Bessel and Riccati-Neumann functions $\hat{j_0}$ and $\hat{\eta_0}$ get simplified as $sin(kr)$ and $-cos(kr)$, respectively and the above equation is written simply as

\begin{equation}
\frac{d \delta_0(k,r)}{dr} = -\frac{V(r)}{k}sin^2[kr+\delta_0(k,r)]
\label{PhaseEqn}
\end{equation}
The function $\delta_{0}(k,r)$ is called phase function. Here, $k = \sqrt{E/(\hbar^2/2\mu)}$ and $\hbar^2/2\mu$ = 41.47 MeVfm$^{2}$. Centre of mass energy $E_{c.m.}$ is related to laboratory energy by $E_{c.m.}=0.5 E_{\ell ab.}$. SPS have been obtained by numerically integrating above equation starting from origin upto asymptotic region using Runge-Kutta (RK) 5th order method \cite{32} with initial condition $\delta_{0}(k,0) = 0$. 
Advantage of PFM is, SPS are directly obtained from potential without recourse to wave-function. So, the Morse function is incorporated into the phase equation and its model parameters are optimised by calling fifth order RK-method in an iterative fashion within an optimisation procedure.
\subsection{Optimisation procedure:}
The procedure utilised for optimisation is broadly as follows:
\begin{enumerate}
\item Model parameters are given certain bounds. For example, both $a_m$ and $r_m$ are chosen to be having values within an interval (0,1).
\item Define a cost function that needs to be minimised. We have chosen mean squared error (\textit{MSE}), between the two data sets, given by
\begin{equation}
MSE = \frac{1}{N} \sum_{i=1}^N \left(\delta_i^{expt} -\delta_i^{sim}\right)^2
\label{MPE}
\end{equation} 
where $\delta_i^{sim}$ are SPS obtained using \textit{PFM} solved via RK-5 method and ($\delta_i^{expt}$) are experimental SPS from  mean energy partial wave analysis data \textit{MEPWAD} of  Granada \cite{33}.
\item Call the optimisation routine to determine the best parameters that fit the experimental data with minimum MSE.
\end{enumerate}
The detailed procedure for obtaining final optimised parameters using TDA is discussed in results and Appendix section.
Once the model parameters are obtained, one can determine the DWF.
\subsection{Deuteron's Ground state wave function:}
To determine deuteron charge and magnetic FFs measured from electron scattering experiments, the knowledge of ground state wavefunction is a basic requirement. The analytical solution for ground state wave function due to Morse potential \cite{26} is given by
\begin{equation}
u_0(z) = N_0 exp(-z/2)z^{\epsilon_0}~~;z(r)=2\lambda e^{-(r-r_m)/a_m}
\label{wfn}
\end{equation}
where  
\begin{equation}
\epsilon_0 = \sqrt{\frac{2\mu E_0 a_m^2}{\hbar^2}}
\label{epsilon}
\end{equation}
and $N_0$ is to be determined from normalisation of Deuteron wave-function (DWF) $\psi_D(r)$. Considering $^3D_1$ wave-function $w_2(r)$ to be proportional to $u_0(r)$ \cite{19,34}, $N_0$ has been determined, such that
\begin{equation}
\int_0^{\infty}\lvert\psi_D(r)\rvert^2r^2dr = \int_0^\infty \left(u_0^2(r) + w^2_2(r)\right)dr = 1
\label{dwf}
\end{equation}
It should be noted that this equation has two unknowns and hence one more condition needs to be utilised to fix them. This is done by choosing one of the static properties of deuteron from experimental data, typically electric quadrupole moment \cite{2}. Considering relativistic effects and deuteron finite size to be negligible the quadrupole moment is given by following expression:
\begin{equation}
Q_D=\frac{1}{20}\int^{\infty}_0 r^2 \bigg(\sqrt{8}u_0(r)w_2(r)- w^2_2(r)\bigg)dr
\label{quad}
\end{equation}
\subsection{Emergent Deuteron properties}
Once the DWF is determined one can determine both static properties and form factors.
\subsubsection{Static properties of Deuteron:}
Static properties like matter radius ($r_{Dm}$), charge radius ($r_{ch}$) and magnetic moment ($\mu_D$) have be determined using following expressions \cite{1,18}:
\begin{equation}
r_{Dm}^2 = \frac{1}{4}\int^{\infty}_0 r^2 \big[u^2_0(r)+ w^2_2(r)\big]dr 
\end{equation}
\begin{equation}
r_{c h}^{2} =  r_{Dm}^{2}+\Delta r_{m}^{2}+r_{p}^{2}+r_{n}^{2}+\frac{3}{4}\left(\frac{\hbar}{m_{p}}\right)^{2}
\end{equation}
and
\begin{equation}
\mu_D =\mu_s - 1.5(\mu_s-0.5)P_d 
\end{equation}
where $r_{p} = 0.862(12) fm$ is charge rms-radius of proton  , ${r_{n}}^{2} = -0.113(5) fm^{2}$ is charge rms-radius of neutron , $\Delta r_{m}^2 = \pm0.01 fm^2$ and  $P_D$ is D-state probability. 
\subsubsection{Deuteron Form Factors:}
For understanding nucleon structure, study of measurable fundamental quantity such as electromagnetic FFs is of paramount importance. The FFs are helpful for describing the spatial variation of the distribution of magnetisation and charge of nucleon within the two nucleon bounded system. Deuteron can't be considered as a point like object. Hence, the elastic electron-deuteron (e-D) scattering process is utilised to probe into the structure of the nucleus to obtain the FFs.\\
In non-relativistic theory, without considering $(v/c)^2$ corrections, the following relations are used for the calculations of em FFs: 
\begin{align}
F_{C}(Q) & =\left[G_{E_p}+G_{E_n}\right] \int_{0}^{\infty}\left[u^{2}_0+w^{2}_2\right] j_{0} d r&\\
F_{Q}(Q)     & =\frac{2}{\zeta} \sqrt{\frac{9}{8}}\left[G_{E_p}+G_{E_n}\right] \int_{0}^{\infty}\left[u_0 w_2-\frac{w^{2}_2}{\sqrt{8}}\right] j_{2} d r\\
F_{M}(Q) & =2\left[G_{M_p}+G_{M_n}\right] \int_{0}^{\infty}\left[\left(u^{2}_0-\frac{w^{2}_2}{2}\right) j_{0}\nonumber+\left(\frac{u_0 w_2}{\sqrt{2}}+\frac{w^{2}_2}{2}\right) j_{2}\right] d r &\\ 
&+\frac{3}{2}\left[G_{E_p}+G_{E n}\right] \int_{0}^{\infty} w^{2}_2\left[j_{0}+j_{2}\right] d r
\end{align}
Where $j_0$ and $j_2$ are the spherical Bessel functions with an argument $(Qr/2)$. While $G_{E_p}$ and $G_{E_n}$ are the proton and neutron isoscalar
electric FF, $G_{M_p}$ and $G_{M_n}$ are the corresponding isoscalar magnetic FF. The factor $\zeta$ is related to 4-momentum transfer $Q$ by
$$
\zeta=\frac{Q^{2}}{4 M_{D}^{2}} ; \quad M_{D}=1875.63 \mathrm{MeV}
$$
Here, the charge FF for the neutron $G_{E_N}$ is assumed to be zero as in \cite{18} and the charge FF for proton $G_{E_P}$ was parametrised using following dipole FF relation
\begin{equation}
G_{E_p} = \frac{1}{(1 + 0.054844Q^2)^2}
\end{equation} 
The magnetic FF for the nucleon is determined using following 
$$G_{M_p}=\mu_p G_{E_p} ~~~~\& ~~~~G_{M_n}=\mu_n G_{E_p}$$
Where $\mu_p=2.7928 ~~\& ~~\mu_n=-1.9130$ are the magnetic moments of proton and neutron having units in nuclear magneton.
We have determined the three FFs directly by integrating the deuteron's analytical wavefunctions. Here, it is to be noted that in non-relativistic limits, the nucleon electric FF contributes to the deuteron charge as well as quadrupole structure, while rest two FFs contribute to the magnetic structure of the deuteron. We can then calculate deuteron structure functions involved in the calculation, A(Q) and B(Q), are related to three electromagnetic (em) FFs due to charge $F_C(Q)$, Quadrupole $F_Q(Q)$ and magnetic $F_M(Q)$, through the following \cite{35,36,37,38,39,40,41,42,43,44}:
\begin{equation}
A(Q)=F^2_C+\frac{8}{9}\zeta^2F^2_Q+\frac{2}{3}\zeta F^2_M
\label{A(Q)}
\end{equation}
\begin{equation}
B(Q)=\frac{4}{3}\zeta (1+\zeta)F^2_M
\label{B(Q)}
\end{equation}
Using further A(Q) and B(Q) yields unpolarised e-D elastic scattering cross section given by following relation \cite{2,18}:
\begin{equation}
\frac{d\sigma}{d\Omega} = \frac{\sigma_{Mott}}{1+\frac{2E}{M_d}sin^2(\frac{\theta_e}{2})}[A(Q)+B(Q)tan^2(\theta_e/2)]
\end{equation}
Where $\sigma_{Mott}$ is Mott cross section given as:

$$\sigma_{Mott}=\alpha^2 E'cos^2(\theta_e/2)/(4E^3 sin^4(\theta_e/2))$$

Here, $Q(fm^{-1})$ is momentum transfer, $\alpha=e^2/4\pi$ = 1/137 is the fine structure-constant, $\theta_e$ is electron scattering angle, $E$ and $E'$ is electron's incident and final scattered energy and $M_D$ is the deuteron mass. 

%


To obtain low energy scattering properties and total scattering cross-section, we need to consider $^1S_0$-singlet state.
\subsection{Singlet S-wave scattering state phase shifts:}
In case of singlet $^1S_0$ the exact relation which we have used for $^1S_0$ SPS calculation is given by Matsumoto and Guerin \cite{26,45} as 
\begin{equation}
\delta_0^{\mathrm{th.}} = -k r_m - \epsilon_0(\gamma + log 2\lambda) +\sum_{i=1}^\infty \left(\frac{\epsilon_0}{i} - tan^{-1}\frac{2\epsilon_0}{i} - tan^{-1}\frac{\epsilon_0}{\lambda + 1/2}\right)
\label{1s0anal}
\end{equation}
Here, $\gamma$ is Euler constant ($\gamma = 0.57721$) and $\epsilon_0$ is given by Eq. \ref{epsilon}.
It is interesting to note that SPS are linearly dependent on $r_m$, but have a non-linear dependence on parameters $V_0$ and $a_m$, for different values of lab energy parameter $k$. 
SPS can be directly obtained using this analytical expression, as an iterative equation, to determine best model parameters that give minimum mean squared error (MSE) with respect to experimental mean energy partial wave analysis data (MEPWAD) \cite{33}.
Alternatively, one can obtain optimised parameters using PFM technique inside the iterative loop. Thus, analytical expression Eq. \ref{1s0anal} provides a good cross-check to validate efficacy of PFM.\\
It is important to remember that the main contribution to total scattering cross-section is from singlet $^1S_0$ state. The results for SPS are shown in \ref{fig1}(a)
\subsection{Partial and Total Cross-sections:}
Partial cross section $\sigma_{\ell}$ can be calculated using
\begin{equation}
\sigma_{\ell}=\frac{4\pi}{k^2}(2\ell+1)sin^2{\delta_{\ell}(k)}
\end{equation} 
and total cross-section \cite{46} is calculated as
\begin{equation}
\sigma(k)=\frac{1}{4}\big(3\sigma_t+\sigma_s\big)
\end{equation}
where $\sigma_t$ and $\sigma_s$ are partial cross-sections for $^3S_1$ and $^1S_0$ respectively.
\subsection{Low energy scattering parameters:}
Low energy scattering parameters i.e. scattering length (\textit{a}) and effective range ($r_e$) are determined from slope and intercept of $k cot(\delta)~ vs~ 0.5~k^2$ relation while the low energy SPS can be obtained using  using following relation \cite{47} for  any potential
\begin{equation}
k cot(\delta)=-\frac{1}{a}+\frac{1}{2}k r^2_e
\end{equation}
A linear regression plot of $k$ vs $kcot(\delta)$  allows for calculation of required scattering parameters $a$ and $r_e$ from intercept and slope respectively.
\section{Results, Analysis and Discussion:}
The experimental MEPWAD for SPS of both $^3S_1$ and $^1S_0$, S-wave channels, have been taken from Arriola \textit{et al.}, (2016), Granada group \cite{33}. This data consists of SPS for lab energies ranging from 1 MeV to 350 MeV, given by the set [1, 5, 10, 25, 50, 100, 150, 200, 250, 300, 350]. Since, scattering parameters depend upon low energy data, it is important to include experimental SPS at low energy. Hence, [E, $\delta$] given by [0.1, 169.32] for $^{3}S_{1}$ and [0.1, 38.43] for $^{1}S_{0}$ data points from Arndt \textit{(Private communication)} have been added. 
\subsection{Overall Data Fitting using Machine Learning paradigm:}
Initially, model parameters are optimised by choosing to minimize MSE for entire data set consisting of 12 points. First, analytical expression for SPS of singlet ($^1S_0$) scattering state, given by Eq. \ref{1s0anal} has been numerically implemented and model parameters are optimised by minimizing MSE between obtained and experimental values. Then, optimisation is performed using 5th order RK method within an iterative loop to minimize MSE. Both procedures have resulted in exactly same values for model parameters, and are shown in Table \ref{scatparam}. This cross-verifies the correctness of PFM.\\
In case of triplet ground state ($^3S_1$), only two parameters $a_m$ and $r_m$ are varied and $V_0$ is calculated via energy constraint Eq. \ref{V0constraint}. The optimised values obtained are shown in Table \ref{scatparam}. The MSE values obtained are $ < 0.1$ and to quantify the performance, we have chosen mean absolute error (MAE) as a measure. The triplet and singlet SPS have been obtained with MAE of 0.35 and 0.70 respectively. 

The uncertainties, $\Delta \delta(E)$, in SPS data at different energies specified in Granada MEPWAD \cite{33} have been utilised to create two extreme data sets. One by adding $\Delta \delta(E)$ to $\delta(E)$ and the other by subtracting $\Delta \delta(E)$ from $\delta(E)$. The model parameters obtained for these two respective sets are:\\
$^3S_1$: [116.040, 0.832, 0.347] $\&$  [112.306, 0.850, 0.354]\\
$^1S_0$: [72.463, 0.891, 0.367] $\&$ [68.521, 0.911, 0.377]\\
These model parameter sets are used to obtain uncertainties in SPS for triplet and singlet states.
While the obtained SPS are utilised to determine low energy scattering parameters and total cross-section, the model parameters give rise to deuteron wave function(DWF) from which various static properties are determined. The DWF also helps in calculation of its various em form factors.

This kind of analysis is akin to data fitting as in MLA, wherein best parameters are obtained by including all available experimental values, at validation stage, to obtain model interaction. One should be aware that there is a good possibility that MLA might lead to over-fitting \cite{25}. Also, optimised parameters could be sensitive to data set. This aspect is being studied.

\begin{table}
\centering

\caption{Optimised parameters for both $^3S_1$ and $^1S_0$ states using MLA and TDA. In later case, parameter values consisting of extreme depths are shown. Scattering length ($a$ in fm) and effective range ($r_e$ in fm) obtained, using SPS determined from these optimised parameters, are shown with experimental values \cite{48} in curly  brackets.} 

\scalebox{0.84}{
\setlength{\tabcolsep}{8pt} 
\renewcommand{\arraystretch}{1.4} 

\begin{tabular}{cccccc} 

\hline\hline

Analysis                & States                   & $[V_0$, $r_m$, $a_m]$     & MAE & $a (fm)$ & $r_e (fm)$   \\ 

\hline

\multirow{2}{*}{MLA}    & $^3S_1$                  & {[}114.153, 0.841, 0.350] & 0.35 & 5.35(1)$\{$5.424(3)$\}$ & 1.75(2)$\{$1.760(5)$\}$   \\

\cline{2-4}

                        & $^1S_0$                  & {[}70.439, 0.901, 0.372]  & 0.70 & -23.37(8)$\{$-23.749(8)$\}$ & 2.42(3)$\{$2.81(5)$\}$\\ 

\hline

\multirow{4}{*}{ TDA  } & \multirow{2}{*}{$^3S_1$} & {[}93.577, 0.843, 0.394]  & 0.4 & \multirow{2}{*}{5.38(2)}&\multirow{2}{*}{1.76(1)}   \\

                        &                          & {[}116.382, 0.843, 0.346] & 1.1    \\ 

\cline{2-4}

                        & \multirow{2}{*}{$^1S_0$} & {[}67.119, 0.897, 0.380]  & 0.8 & \multirow{2}{*}{-22.54(1.30)}&\multirow{2}{*}{2.41(1)} \\

                        \vspace{0.05cm}

                        &                          & {[}74.976, 0.897, 0.361]  & 0.9   \\

\hline

\end{tabular}
}
\label{scatparam}
\end{table}
\subsubsection{Scattering phase shifts and cross-section:}
\begin{figure*}[htp]
\centering
{\includegraphics[scale=0.42, angle=-90]{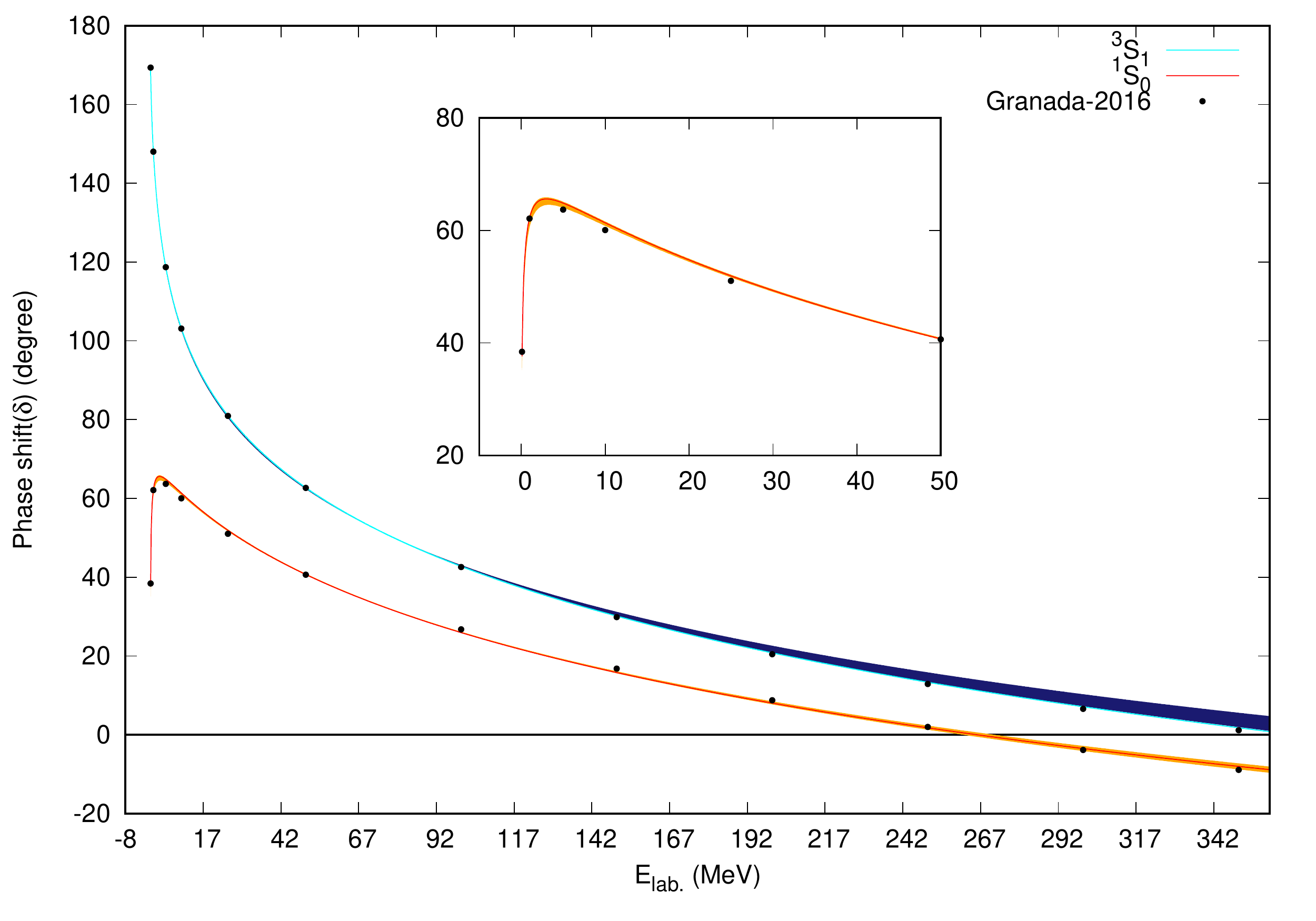}}
{\includegraphics[scale=0.42, angle=-90]{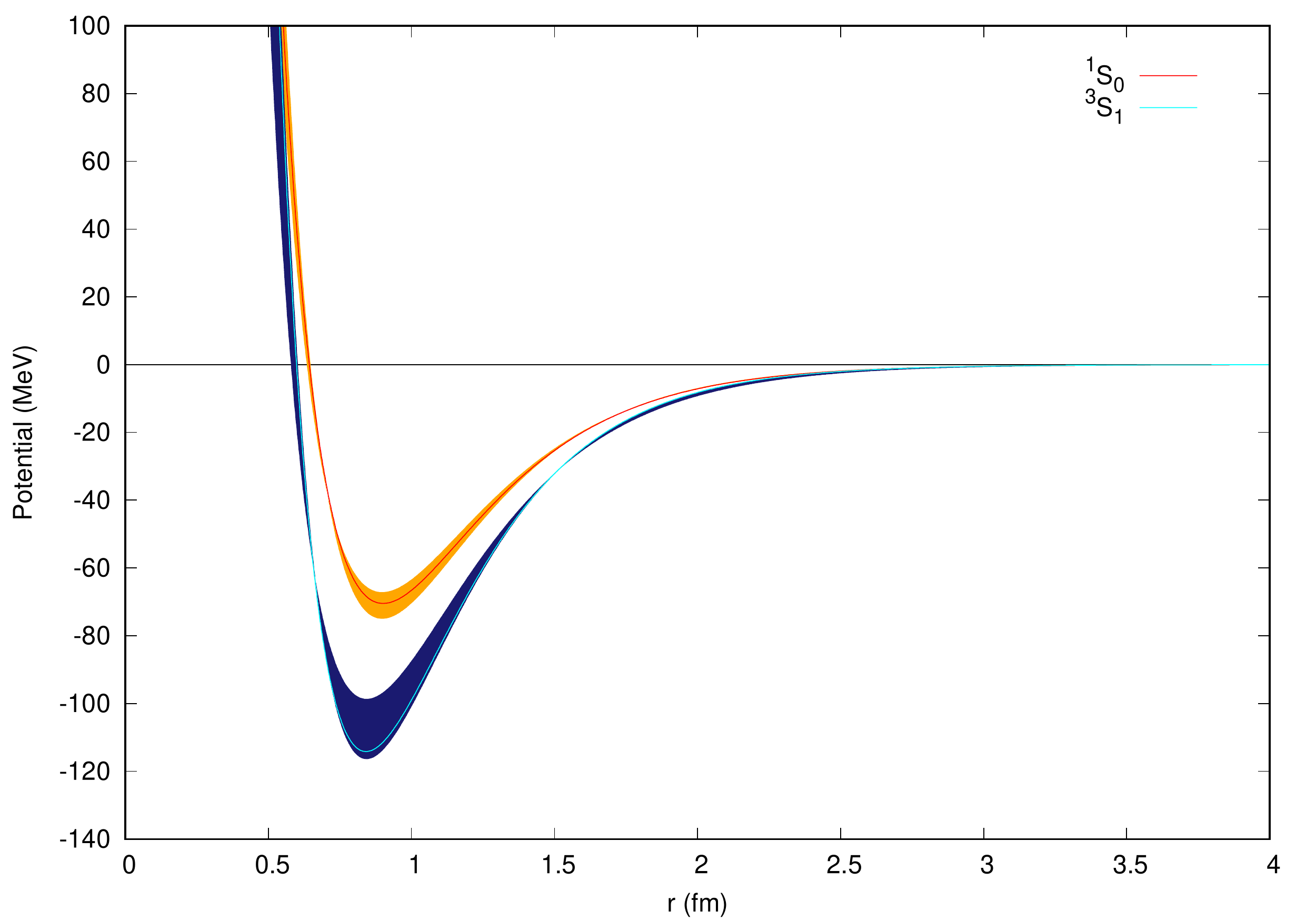}}
\caption{(a) Triplet and singlet scattering phase shifts at lab energies compared with experimental MEPWAD \cite{44} (b) Interaction potentials obtained using TDA. The bold lines are best fits obtained using MLA.} 
\label{fig1}
\end{figure*}

The SPS for $^{3}S_{1}$ and $^{1}S_{0}$, obtained from MLA, are shown in fig. \ref{fig1}(a) using a bold line. The corresponding interaction potentials are shown in fig. \ref{fig1}(b). The variation in SPS and certain width seen in figs.\ref{fig1}(a) and \ref{fig1}(b) respectively are due to uncertainties calculated in model parameters using traditional data analysis (TDA) discussed in Sec 3.2. The uncertainties due to extreme data sets discussed above are within those from TDA and hence are not separately shown.

\begin{figure*}[htp]
\centering
{\includegraphics[scale=0.4, angle=-90]{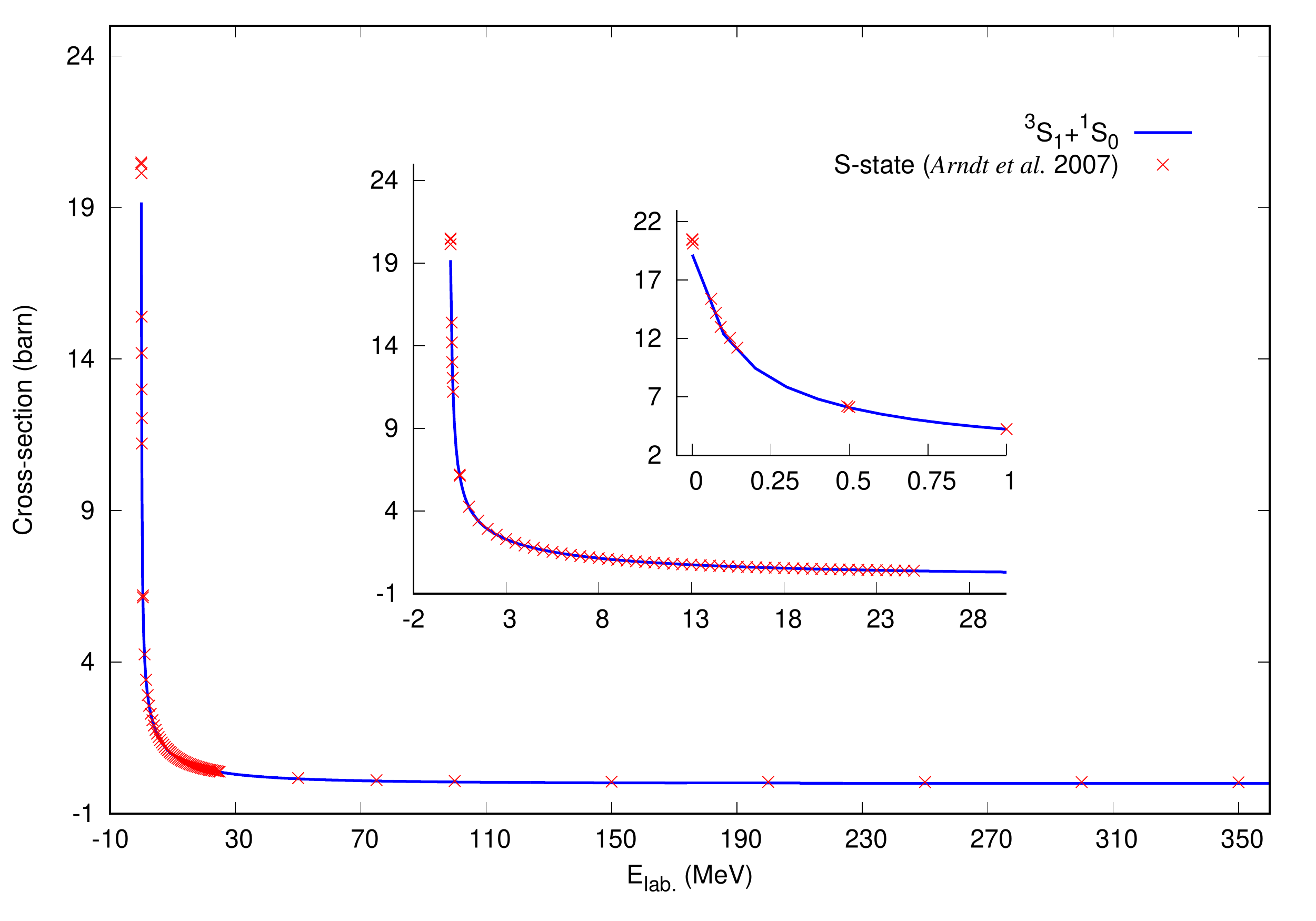}}
\caption{Total scattering cross-section plotted against lab energies. The experimental values are from Arndt \cite{49}.} 
\label{tcs}
\end{figure*}

Fig. \ref{tcs} shows total cross-sections calculated at all energies from 0.132 KeV to 350 MeV with appropriate insets to emphasize excellent match with experimental data. The calculated (experimental) cross section at E = 0.132  KeV was found to be 19.171 $\pm$ 1.280 (20.491 $\pm$ 0.014) barn. SPS for all higher $\ell$-channels are also found to be matching well with experimental data (to be communicated separately) and their contributions will further enhance the obtained value of total scattering cross-section.
 
\subsubsection{Low energy scattering and static properties of Deuteron:}
\begin{figure*}[htp]
\centering
{\includegraphics[scale=0.4, angle=-90]{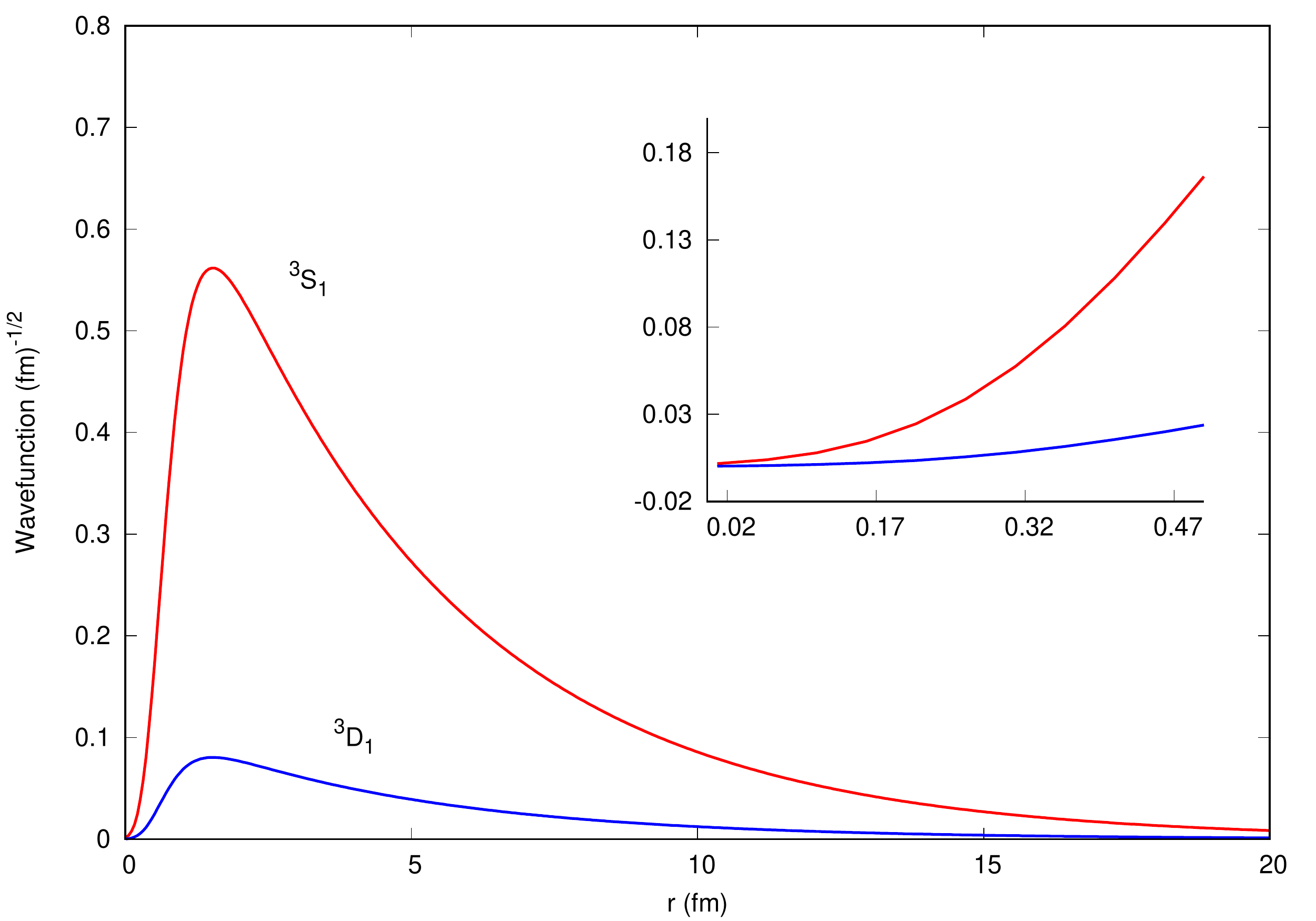}}
\caption{Analytical Deuteron wave function for $^3S_1$ and $^3D_1$ states. Inset shows the variation of wavefunctions closer to the origin.} 
\label{wf}
\end{figure*}

Low energy parameters, scattering length (\textit{a}) and effective range ($r_e$) have been obtained for both S-waves by plotting graphs of $kcot(\delta)$ vs $k$. The slope and intercept give rise to $a$ and $r_e$. The results are compared alongside experimental ones, given in brackets, in upper half of Table \ref{scatparam}. Once again the extreme data sets for model parameters were utilised to present uncertainties for low energy properties in Table \ref{scatparam}.\\
The $^3S_1$ ground state wave-function $u_0(r)$ has been determined by substituting the model parameters in Eq. \ref{wfn} and is shown in fig. \ref{wf}. The $^3D_1$ wave-function $w_2(r)$, also shown in fig. \ref{wf}, has been determined so as to ensure normalisation and  correct electric quadrupole moment value of 0.2589 fm$^2$. 
Due to the repulsive nuclear core, the wavefunctions can be seen to be dropping sharply near to the origin, while the peaks for $u_0(r)$ and $w_2(r)$ occur in the intermediate range (1 $\leq$ r $\leq$ 2 fm)

The average values of S-state and D-state probabilities, $P_S$ and $P_D$, are obtained as 98\% and 2\% respectively. Rest of the static properties, magnetic moment, matter and charge radii are determined, along with their uncertainties, and are presented in Table \ref{statprop}. 

\subsubsection{Deuteron Form Factors:}
\begin{figure*}[htp]
\centering
{\includegraphics[scale=0.56,angle=0]{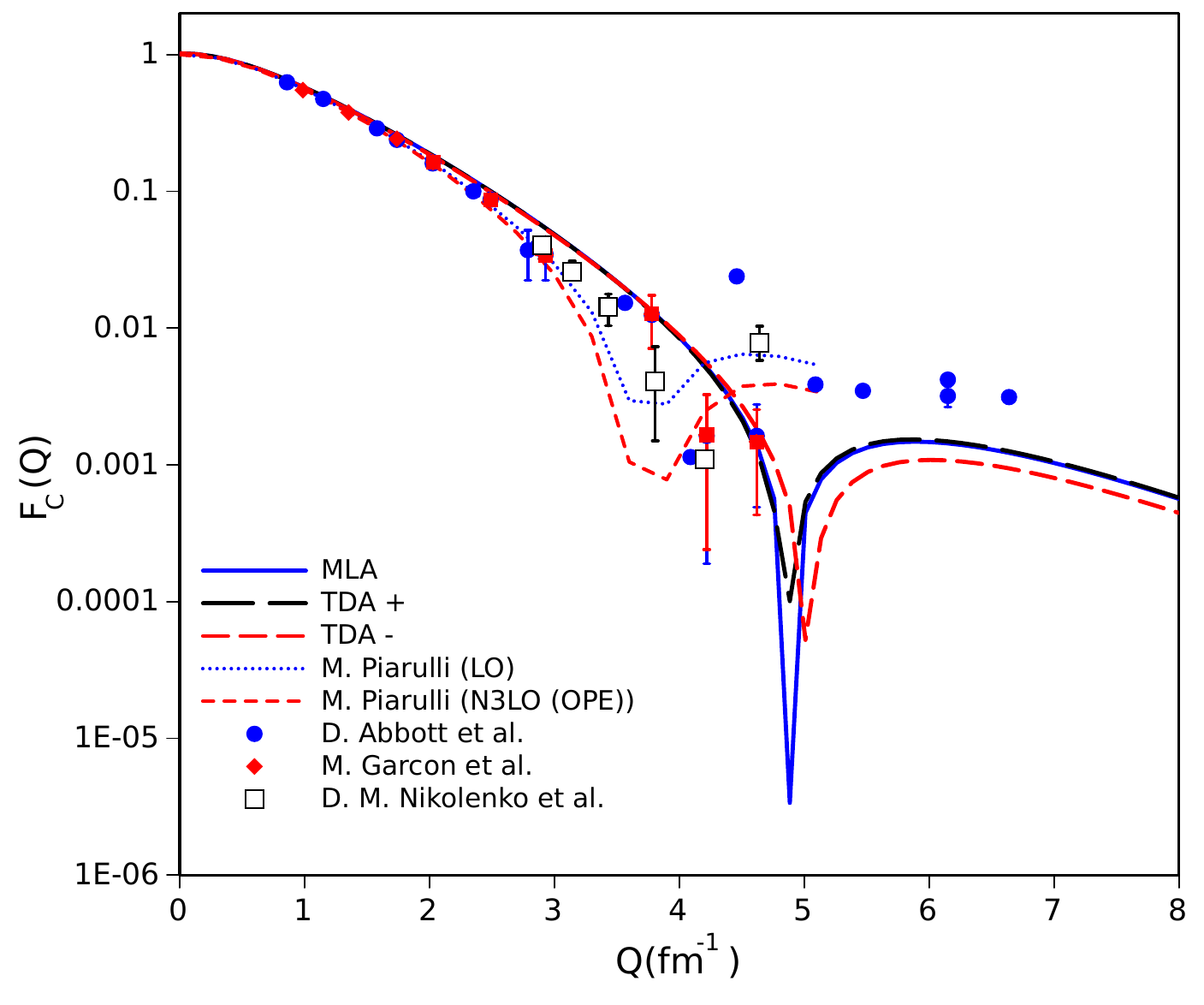}}
{\includegraphics[scale=0.56,angle=0]{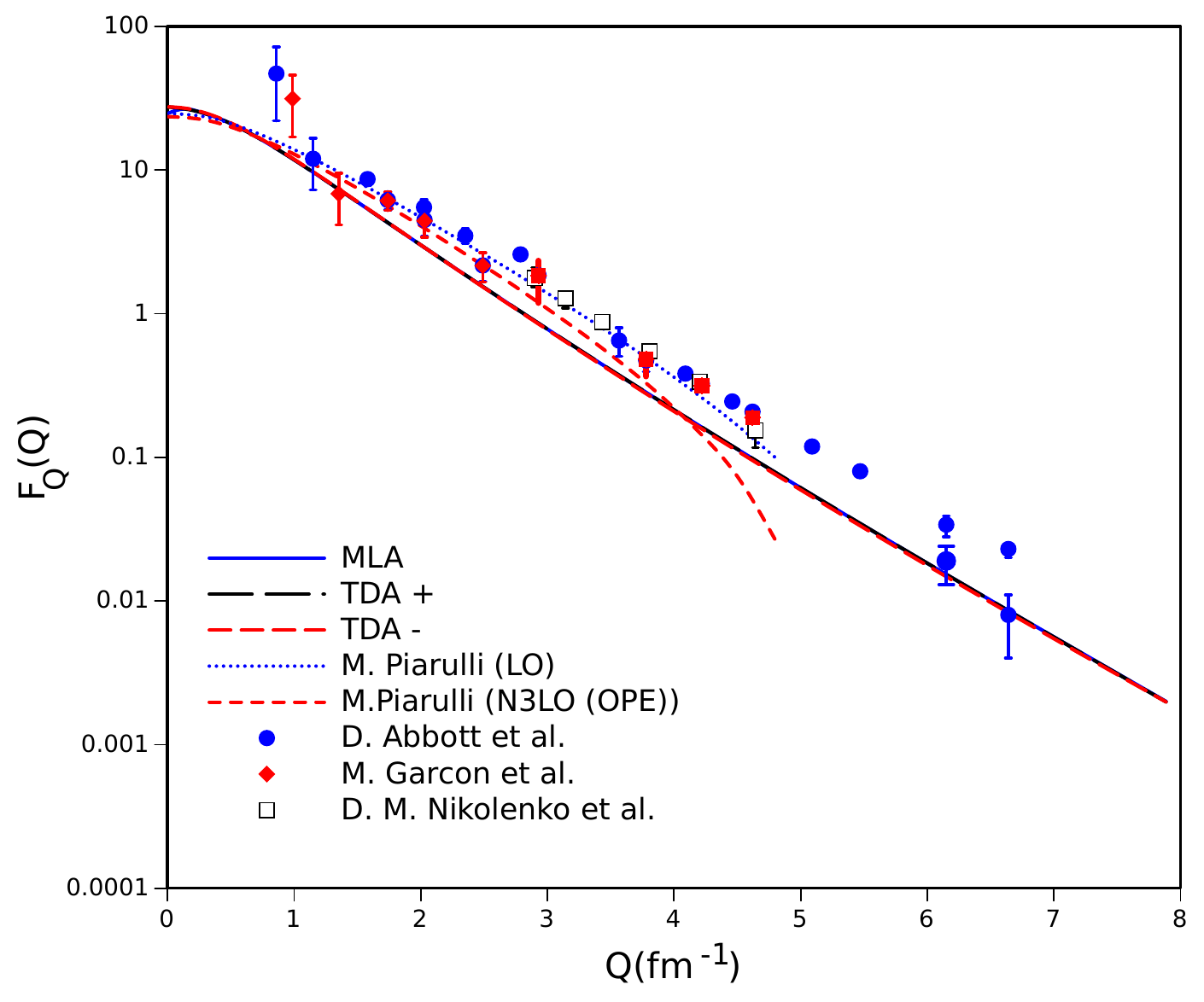}}
{\includegraphics[scale=0.56,angle=0]{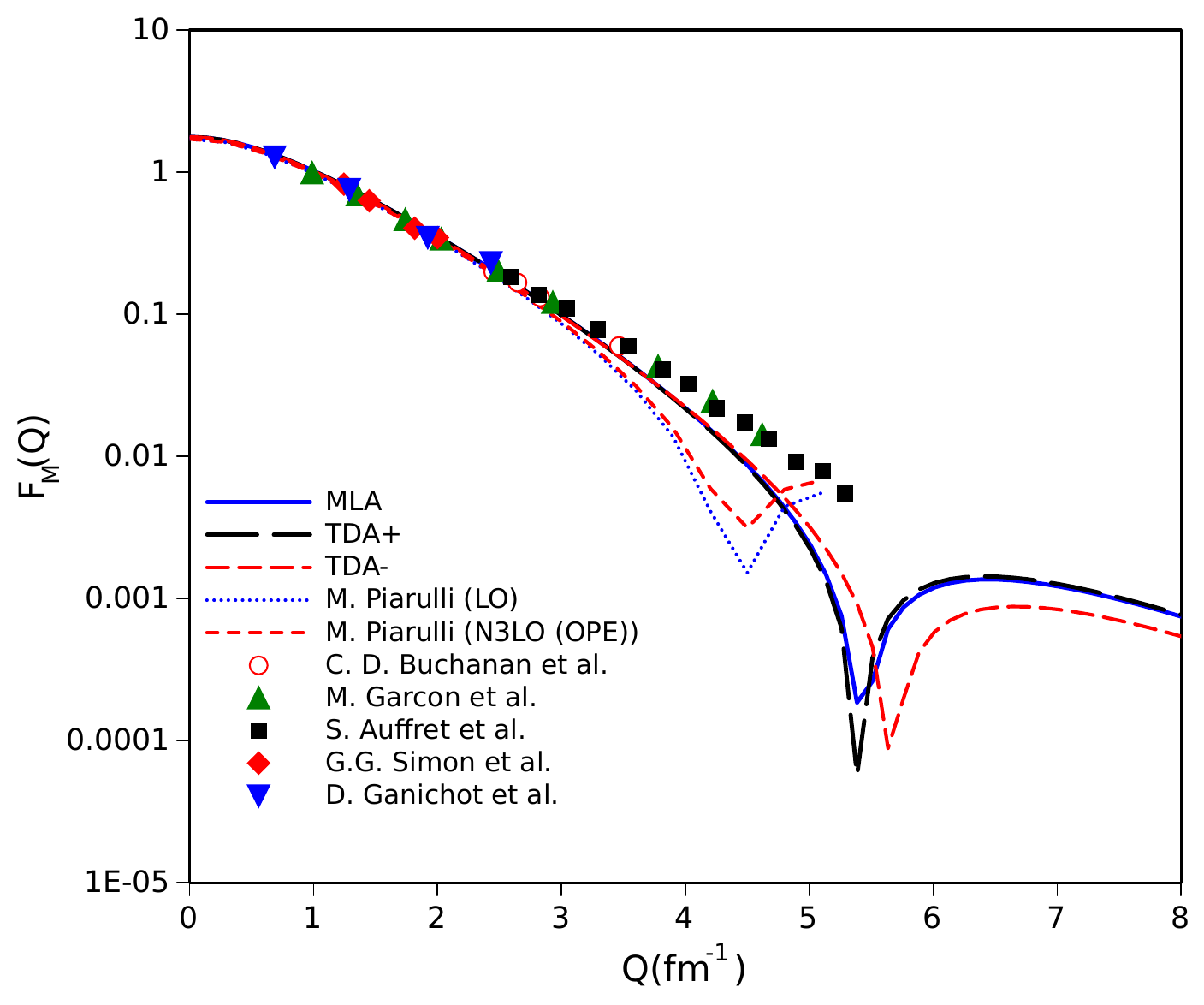}}
\caption{Deuteron form factors $F_C, F_Q$ and $F_M$ as a function of Q. Experimental data are taken from different experimental works \cite{35,36,37,38,39,40,41,42,43,44}. Leading order (LO) and N3LO have been taken from \cite{50} for comparison.} 
\label{FFs}
\end{figure*}

\begin{figure*}[htp]
\centering
{\includegraphics[scale=0.6, angle=0]{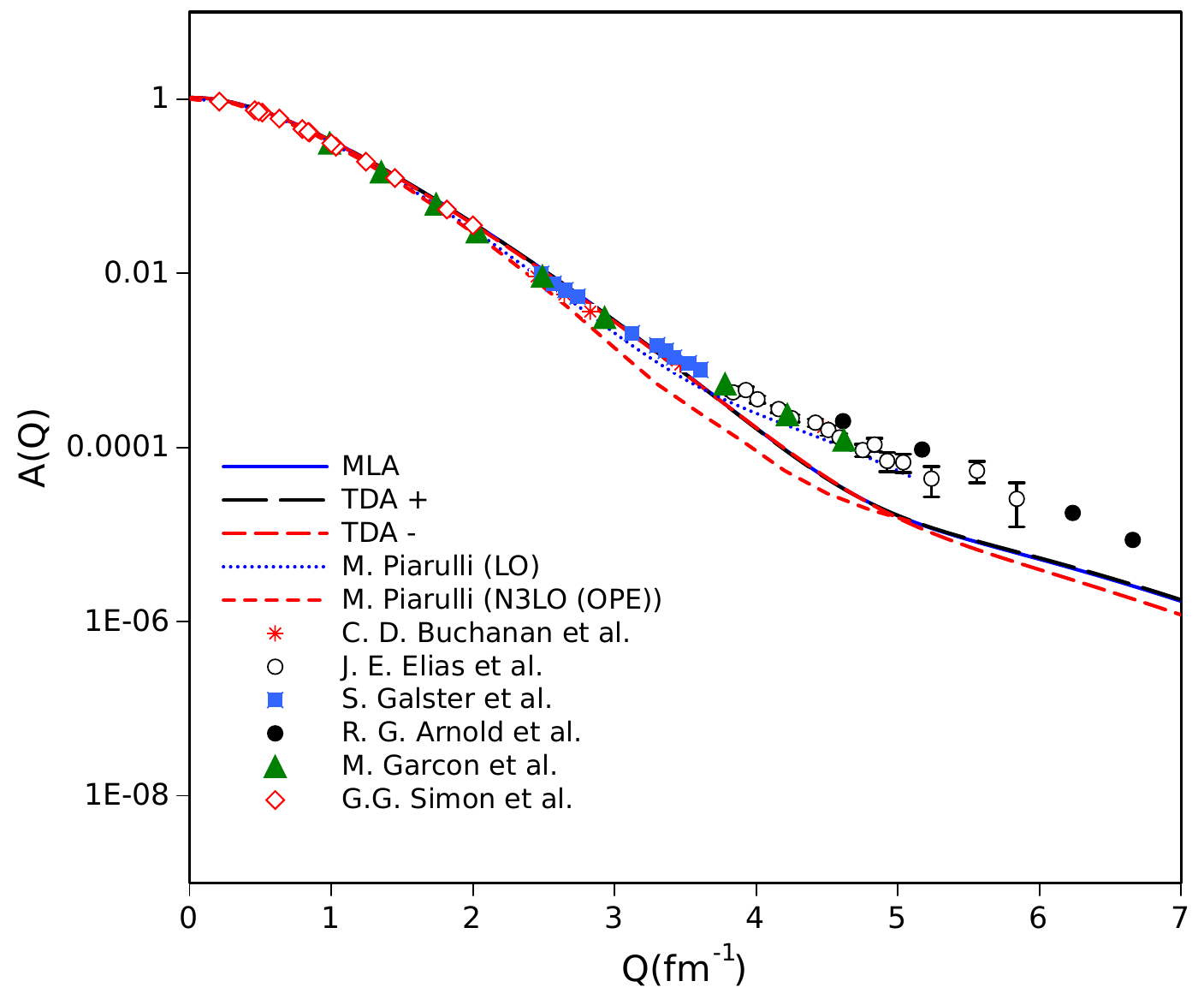}}
{\includegraphics[scale=0.6, angle=0]{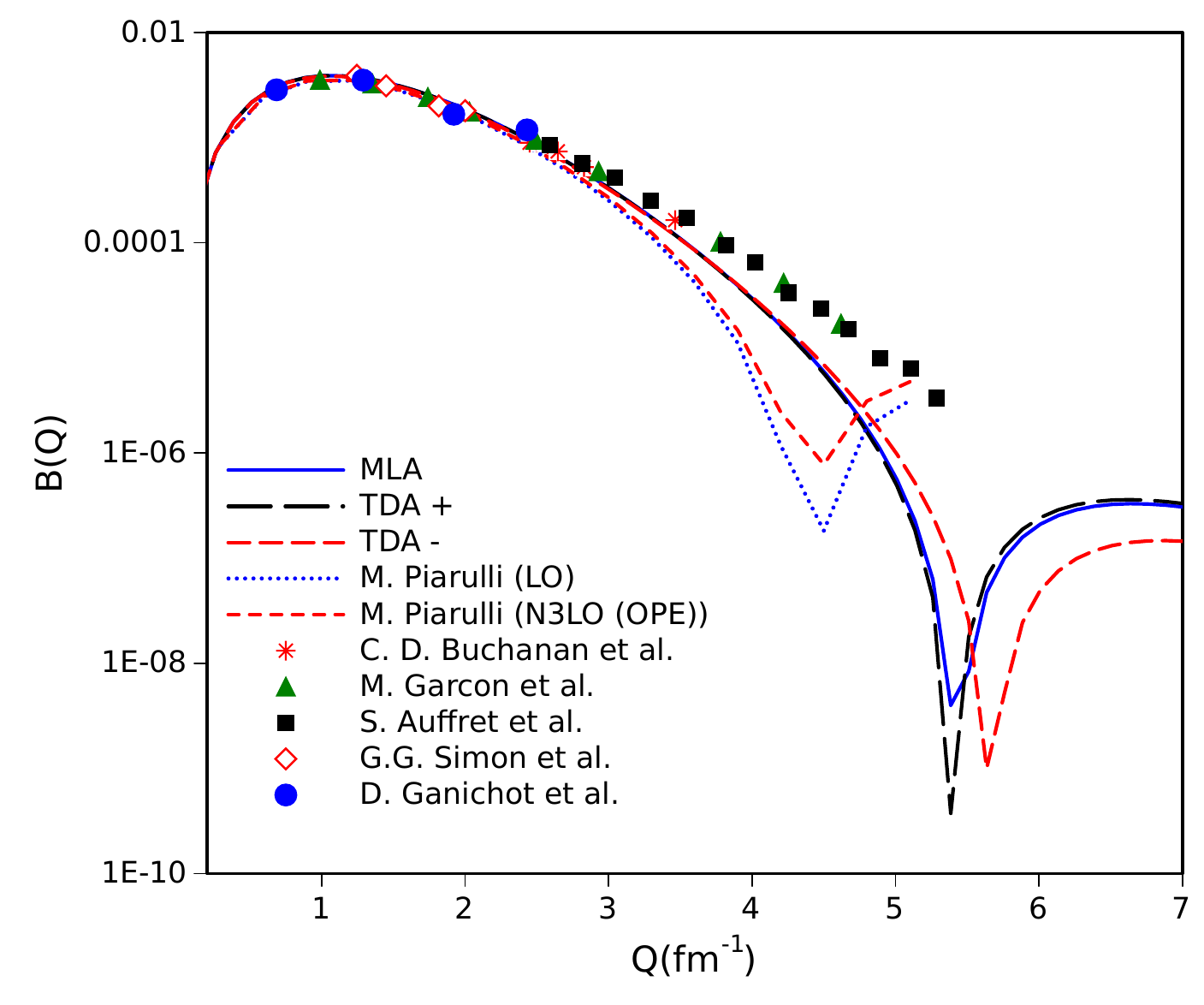}}
\caption{Deuteron electric $A(Q)$ and magnetic $B(Q)$ structure functions variation with \textit{Q}($fm^{-1}$). Experimental values are taken from \cite{35,36,37},\cite{39,40,41,42} and \cite{44}. Leading order (LO) and N3LO have been taken from \cite{50} for comparison.}
\label{fig2}
\end{figure*}

The analytical wavefunctions $u_0(r)$ and $w_2(r)$ have been directly used in the integrals, in Eqs. $20-22$, to determine the em FFs $F_C(Q)$, $F_Q(Q)$ and $F_M(Q)$ respectively.  The integral calculations are performed using symbolic python. These are plotted in fig. \ref{FFs}. One can see good match with experimental data \cite{35,36,37,38,39,40,41,42,43,44} for lower momentum transfer, Q values, upto 3-4 fm$^{-1}$ after which our values slowly deviate from expected. 
As Q $\rightarrow$ 0 (static moments), the values of three FFs obtained(experiment\cite{2}) are as follows:
\begin{align*} 
F_C(Q \rightarrow 0) &=  1.0205 \textbf{(1)} \\ 
F_M(Q \rightarrow 0) &=  1.7714  \textbf{(1.7148)}\\
F_Q(Q \rightarrow 0) &=  24.9724 - 27.5613  \textbf{(25.83)} 
\end{align*}
and can be seen to be in good agreement.\\
In case of $F_C(Q)$, the experimental data is available from nearly 1 fm$^{-1}$ to close to 7 fm$^{-1}$ from different papers. Both Abbot's compiled data \cite{38} (0.86-6.64 fm$^{-1}$) and Nikolenko et al. \cite{43} (Q = 2.9-4.6 fm$^{-1}$) indicate an upward going trend. While Garcon et al. \cite{44} (Q = 0.988-4.62 fm$^{-1}$) does not capture the up going trend beyond 4 fm$^{-1}$, both LO and N3LO \cite{50} capture the experimental trend to occur just before 4 fm$^{-1}$. The former matches Nikolenko trend and the later shows closeness to Abbot data. Our analysis fig. \ref{FFs} shows the upward trend to be occurring closer to 5 fm$^{-1}$ and our data beyond 5 fm$^{-1}$ falls below the experimental values. \\
In case of $F_Q(Q)$, all the available experimental data \cite{38}, \cite{43,44} have similar trend. Both LO and N3LO match our values of F(Q) at Q $\approx$ 0. LO calculations match experimental data for Q values upto 4.62 fm$^{-1}$. For values beyond, the trend of LO is downwards as compared to experimental data of Abbot et al. \cite{38}. On the other hand, N3LO calculations are slightly below experimental data from Nikolenko et al. and Abbot et al., and capture Garcon data better for Q value upto 3.78 fm$^{-1}$. Beyond this N3LO bends farther away from both LO calculations as well as experimental data. Our analysis shown in fig. \ref{FFs} lies below the experimental data for Q values upto 4.62 fm$^{-1}$ but correctly obtains the values for Q = 6.15 fm$^{-1}$ \& 6.64 fm$^{-1}$.\\
In case of $F_M(Q)$, Garcon et al. \cite{44} (Q = 0.988-4.62 fm$^{-1}$) captures in essence the trends from both Ganichot et al. \cite{40} (Q = 0.68-2.43 fm$^{-1}$) and Auffret et al. \cite{37} (Q = 2.59-5.28 fm$^{-1}$). While LO and N3LO suggest a dip at around 4.5 fm$^{-1}$, the experimental data does not show such trend. Our analysis fig. \ref{FFs} correctly matches upto 3.5 fm$^{-1}$ and then slowly tends to go farther as Q increases and indicates a dip at around 5.5 fm$^{-1}$.\\
Next, these three em FFs are in turn used in Eq. \ref{A(Q)} \& \ref{B(Q)} to obtain the structure form factors A(Q) and B(Q), shown in fig.\ref{fig2}. Deuteron's electric $A_Q$ and magnetic $B_Q$  structure functions are in quite good match with the experimental data \cite{35,36,37},\cite{39,40,41,42} and \cite{44}.\\
In case of A(Q) the experimental data covers from around 0.01 to 10 fm$^{-1}$ by various experimental works, from Simon et al. \cite{39} (Q = 1.24-2 fm$^{-1}$), Garcon et al. \cite{44} (Q=0.988-4.62 fm$^{-1}$), Galster et al. \cite{44} (Q = 2.48-3.61 fm$^{-1}$), Elias et al. \cite{35} (Q = 3.83-5.84 fm$^{-1}$) and finally Arnold et al. \cite{42} (Q = 4.61-10.04 fm$^{-1}$). LO calculations catch the trend from experiment all the way upto 5 fm$^{-1}$ and may correctly catch higher values on extension. N3LO on other hand matches experimental data only upto 2 fm$^{-1}$ and then falls short increasingly with increasing Q.\\
The experimental data for B(Q) is from Ganichot et al. \cite{40} (Q = 0.68-2.43 fm$^{-1}$) , Garcon et al. \cite{44} (Q = 0.988-4.62), Simon et al. \cite{39} (Q = 1.24-2 fm$^{-1}$)  and Auffret et al. \cite{37} (Q = 2.59-5.28 fm$^{-1}$).  All the data more or less shows similar trend. While LO and N3LO correctly follow the values upto 3 fm$^{-1}$, for points beyond they are way lower. Also, they tend to predict a dip at about 4.5 fm$^{-1}$. On the other hand, our analysis shown in fig. \ref{fig2} also matches with experimental values upto 3 fm$^{-1}$ but is more closer for points beyond as well. It predicts a dip at around 5.5 fm$^{-1}$ after which B(Q) value is increased.\\

\begin{table}
\centering
\caption{Static properties for Deuteron calculated using Morse potential in comparison with experimental values taken from \cite{2} and others \cite{18,19}. Numerical value given with * is calculated using $P_D$ given in \cite{19}}
\begin{tabular}{ccc|ccc} 
\hline
\multirow{2}{*}{Quantity} & \multirow{2}{*}{\begin{tabular}[c]{@{}c@{}}Expt.\\\cite{2}\end{tabular}} & \multicolumn{2}{c}{Our} & \multirow{2}{*}{\cite{18}} & \multirow{2}{*}{\cite{19}}  \\ 
\cline{3-4}
                          &                                                                                            & TDA        & MLA        &                                             &                                              \\ 
\hline
$\mu_D(\mu_N)$            & 0.8574                                                                                     & 0.8687(1)  & 0.8683(1)  & 0.8519(72)                                  & 0.8690$^*$                                            \\
$r_{Dm}(fm)$                & 1.975(3)                                                                                   & 1.9537(39) & 1.9285(44) & 1.95320(475)                                & 1.97507(78)                                  \\
$r_{ch}(fm)$                & 2.130(10)                                                                                  & 2.1088(36) & 2.1037(41) & 2.1354(9)                                   & 2.12562(78)                                  \\
\hline
\end{tabular}
\label{statprop}
\end{table}

\subsection{Data Analysis as in Physics Modeling:}
In principle, for modeling in physics context, one would expect that number of data points to be chosen for optimisation should be equal to number of model parameters. The obtained parameters must be able to explain rest of the data reasonably well. Then, a question arises as to, what would be right way to choose two(three) data points from among the twelve available for triplet(singlet) states. This would result in a total of $^{12}C_2 (= 66)$ and $^{12}C_3 (= 220)$ combinations, for triplet and singlet states, respectively. Initially, we have obtained optimised parameters for each of these available combinations and then carefully analyzed the results. Here, are some important observations:
\begin{enumerate}
\item Depth of potential is energy dependent. That is, data points from low energy region [0.1,  25] have resulted in lower $V_0$ values as compared to those from higher energy regions, [200, 350].
\item The model has good predictive power for interpolated data points but errors increase due to extrapolation, especially at far away points. 
\end{enumerate}
For instance, considering data points from low energy region [0.1, 10] have resulted in better prediction of SPS for immediate data points in range [25, 150] as compared to those in high energy region [200, 350] where the SPS obtained had larger errors. Similarly, considering 3 data points in high energy region [200, 350] for optimisationhave resulted in poor accuracy in scattering parameter values reflecting that SPS of low energy region, important for calculating scattering length and effective range, are not determined to good accuracy. \\
Based on these observations, we have deduced that data points consisting of end points 0.1 and 350 along with an intermediate data point preferably chosen from [25,150] range would give best results. To accommodate more choices, mean absolute error (MAE) was utilised as a quantitative measure. After carefully analysing the results, at various stages of our calculations for the possible combinations, we have applied the following criteria:\\
In the first step, for fixing $r_m$ value, we have considered those combinations for which MAE $\leq$ 2. \\ 
Then, in second step, scattering and static properties as well as scattering phase shifts were obtained for each of the combinations, with MAE $\leq$ 1, and their averages and standard deviations are determined and tabulated.\\ 
This is traditional data analysis (TDA), typically expected, in validation stage of a model.
\subsubsection{Data Analysis of $^3S_1$ state:}
Keeping in mind that, ground state energy is retained, in case of $^3S_1$ through Eq \ref{V0constraint}, it is expected that one should consider equilibrium value for $r_m$ while determining SPS. It was observed that out of 66 combinations of $a_m$ and $r_m$, 64 of them have resulted in MAE $\leq$ 2, which gave $r_m = 0.843 \pm 0.013$ fm. Once $r_m$ value is fixed, there is only one parameter $a_m$ that needs to be determined. Hence, only one energy data point is required to determine $a_m$. That is, a total of 12 values will be obtained for $a_m$ from which corresponding $V_0$ shall be determined to required accuracy such that energy is retained to 6 decimal places. The resultant model parameters shown in Table \ref{scatparam}(3$^{\text{rd}}$ column) correspond to combinations giving rise to two extreme potential depths. It was found that all the 12 combinations resulted in MAE $\leq$ 1. Hence, all of them are considered for determining final properties. Electric quadrupole moment ($Q_D$) is retained in each of the calculations to obtain appropriate $w_2(r)$. Then, magnetic moment ($\mu_D$) and matter radius ($r_{Dm}$) are determined for each combination. The averages along with uncertainties given in Table \ref{statprop} are found to be very close to expected experimental values and comparable to those obtained using realistic precision potentials. The $^3S_1$ SPS with uncertainties and corresponding interaction potential, with shaded regions covering all possible depths  are shown in figs. \ref{fig1}(a) and \ref{fig1}(b) respectively. Similarly, various deuteron form factors have also been determined and are shown in figs. \ref{FFs} and \ref{fig2} with legends as TDA+ and TDA-. 
\subsubsection{Data Analysis of $^1S_0$ state:}
Since, $^1S_0$ SPS are linearly dependent on $r_m$, it is suggestive that it can be replaced by its average value to determine further variations in $V_0$ and $a_m$. To obtain this average, we have considered $r_m$ values from 130 combinations with MAE $\leq$ 2 and obtained $r_m = 0.897 \pm 0.036$ fm. Fixing $r_m$ leaves us with only two parameters to be redetermined from 66 combinations. Only 14 combinations, with MAE $\leq 1$, are utilised for determining scattering length and effective range (See Table \ref{scatparam}, bottom part). Once again, figs. \ref{fig1}(a) and \ref{fig1}(b) show SPS and interaction potentials, for $^1S_0$, along with error margins as shaded regions.

\section{Conclusions:}
In this work, we have obtained an analytical ground state wavefunction for Deuteron by utilizing the analytical properties of Morse potential, that has been constructed using inverse approach. The phase function method was utilised for determining scattering phase shifts at different lab energies for which experimental data are available. Model parameters, two for triplet and three for singlet states, are obtained using machine learning data fitting algorithms as well as traditional data analysis by minimizing mean squared error for mean energy partial wave analysis scattering phase shifts data. Low energy scattering parameters determined for both S-waves are matching with experimental values. $^1S_0$ scattering phase shifts are obtained exactly using analytical formula. The total cross-section due to S-waves contribution has been calculated to be 19.17 barn as compared to experimental value of 20.49 barn at 0.132 MeV lab energy. Both ground state energy and deuteron wave function are obtained analytically. Static properties obtained are close to experimental values.  utilizing the wavefunction, we calculated the FFs which are in good match with those obtained using different experiments.   

\bibliographystyle{elsarticle-num-names} 

\bibliography{cas-refs}
\section{Appendix}
In this section, we have provided all the data analysis tables for sake of clarity in procedure followed.
\subsection{Triplet state analysis:}
There are a total of 12 experimental data points. As discussed in the main paper, to obtain two model parameters of $^3S_1$ state, there are a total of $^{12}C_2 = 66$ possible combinations. The model parameters are determined by choosing two lab energies at a time and minimising the mean squared error. Then, the SPS were obtained at remaining 10 energies from the data set and overall mean absolute error (MAE) is determined. The data has been sorted with ascending values of MAE and is presented in Table \ref{3S1combinations}.
\label{sec:sample:appendix}

\begin{longtable}{ccccccc}

\caption{\textbf{$^3S_1$ state:} Model parameters for 64 combinations, each with two lab energies and obtained by minimising MSE. The overall MAE is determined by obtaining SPS for remaining experimental data points. The data is sorted with respect to MAE in ascending order.}\\
\hline
\textbf{Sr. No.} & \multirow{2}{*}{\begin{tabular}[c]{@{}c@{}}$E_1$\\(MeV)\end{tabular}}  & \multirow{2}{*}{\begin{tabular}[c]{@{}c@{}}$E_2$\\(MeV)\end{tabular}}  & \multirow{2}{*}{\begin{tabular}[c]{@{}c@{}}$V_0$\\(MeV)\end{tabular}}  & \multirow{2}{*}{\begin{tabular}[c]{@{}c@{}}$r_m$\\(fm)\end{tabular}}  & \multirow{2}{*}{\begin{tabular}[c]{@{}c@{}}$a_m$\\(fm)\end{tabular}}   & \multirow{2}{*}{\begin{tabular}[c]{@{}c@{}}\textbf{\small{Overall}}\\\textbf{\small{MAE}}\end{tabular}}   \\  
\\
\hline
\endfirsthead
\multicolumn{7}{c}%
{\tablename\ \thetable\ -- \textit{Continued from previous page}} \\
\hline
\textbf{Sr. No.} & \multirow{2}{*}{\begin{tabular}[c]{@{}c@{}}$E_1$\\(MeV)\end{tabular}}  & \multirow{2}{*}{\begin{tabular}[c]{@{}c@{}}$E_2$\\(MeV)\end{tabular}}  & \multirow{2}{*}{\begin{tabular}[c]{@{}c@{}}$V_0$\\(MeV)\end{tabular}}  & \multirow{2}{*}{\begin{tabular}[c]{@{}c@{}}$r_m$\\(fm)\end{tabular}}  & \multirow{2}{*}{\begin{tabular}[c]{@{}c@{}}$a_m$\\(fm)\end{tabular}}   & \multirow{2}{*}{\begin{tabular}[c]{@{}c@{}}\textbf{\small{Overall}}\\\textbf{\small{MAE}}\end{tabular}}   \\ 
\\
\hline
\endhead
\hline \multicolumn{7}{r}{\textit{Continued on next page}} \\
\endfoot
\hline
\endlastfoot
1	&50	&250	&111.266856	&0.842113	&0.355523		&0.343075	\\
2	&50	&300	&113.765975	&0.842744	&0.350958		&0.347999	\\
3	&100	&250	&118.160348	&0.834258	&0.343322		&0.366430	\\
4	&100	&300	&121.142160	&0.833568	&0.338404		&0.371729	\\
5	&25	&250	&106.206812	&0.848677	&0.365310		&0.378720	\\
6	&25	&300	&108.435228	&0.850302	&0.360906		&0.391483	\\
7	&100	&200	&114.857781	&0.835119	&0.349016		&0.404778	\\
8	&50	&200	&108.514670	&0.841394	&0.360752		&0.405273	\\
9	&50	&350	&115.985987	&0.843294	&0.347040		&0.414131	\\
10	&100	&350	&123.749333	&0.833029	&0.334266		&0.418988	\\
11	&150	&250	&123.820300	&0.828617	&0.334155		&0.427389	\\
12	&150	&200	&119.988203	&0.830672	&0.340284		&0.434592	\\
13	&150	&300	&127.225352	&0.826957	&0.328967		&0.444056	\\
14	&25	&200	&103.759467	&0.846741	&0.370328		&0.450475	\\
15	&0.1	&250	&101.732907	&0.855129	&0.374635		&0.461311	\\
16	&10	&250	&101.457497	&0.855548	&0.375232		&0.466876	\\
17	&0.1	&10	&102.291602	&0.856723	&0.373433		&0.478300	\\
18	&25	&350	&110.429393	&0.851669	&0.357091		&0.479034	\\
19	&100	&150	&111.287630	&0.836170	&0.355484		&0.481976	\\
20	&150	&350	&130.137021	&0.825652	&0.324709		&0.486032	\\
21	&10	&300	&103.423572	&0.858233	&0.371032		&0.493101	\\
22	&0.1	&300	&103.087498	&0.858796	&0.371740		&0.503281	\\
23	&200	&250	&128.974759	&0.824029	&0.326390		&0.504515	\\
24	&0.1	&200	&100.432380	&0.850850	&0.377476		&0.517546	\\
25	&5	&250	&99.257965	&0.858988	&0.380094		&0.532574	\\
26	&50	&150	&105.505377	&0.840559	&0.366728		&0.536353	\\
27	&200	&300	&132.690407	&0.821650	&0.321103		&0.540245	\\
28	&10	&200	&99.341801	&0.852269	&0.379906		&0.541456	\\
29	&5	&300	&101.023747	&0.862347	&0.376176		&0.569436	\\
30	&200	&350	&135.785874	&0.819813	&0.316882		&0.585140	\\
31	&1	&250	&97.468905	&0.861917	&0.384185		&0.601450	\\
32	&5	&200	&97.414856	&0.854866	&0.384310		&0.602077	\\
33	&10	&350	&105.213478	&0.860447	&0.367323		&0.606850	\\
34	&25	&150	&101.073042	&0.844377	&0.376069		&0.614751	\\
35	&0.1	&350	&104.415753	&0.861876	&0.368963		&0.634760	\\
36	&250	&300	&137.452012	&0.817454	&0.314674		&0.638710	\\
37	&1	&200	&96.100208	&0.856706	&0.387400		&0.654530	\\
38	&1	&300	&98.876825	&0.866218	&0.380955		&0.657098	\\
39	&0.1	&150	&99.269504	&0.846052	&0.380068		&0.664225	\\
40	&250	&350	&140.600289	&0.815282	&0.310622		&0.691583	\\
41	&5	&350	&102.667292	&0.865094	&0.372631		&0.699167	\\
42	&10	&150	&97.091043	&0.848173	&0.385065		&0.728276	\\
43	&50	&100	&102.044546	&0.839500	&0.373963		&0.745526	\\
44	&300	&350	&144.473700	&0.811904	&0.305835		&0.786923	\\
45	&5	&150	&95.544952	&0.849747	&0.388727		&0.788417	\\
46	&1	&350	&100.256305	&0.869734	&0.377865		&0.809464	\\
47	&1	&150	&94.851972	&0.850471	&0.390400		&0.821075	\\
48	&0.1	&25	&98.433360	&0.841721	&0.381964		&0.850510	\\
49	&0.1	&100	&98.321794	&0.841081	&0.382219		&0.879760	\\
50	&25	&100	&97.986161	&0.841230	&0.382989		&0.892571	\\
51	&10	&100	&94.671293	&0.842749	&0.390840		&1.048340	\\
52	&0.1	&50	&97.853099	&0.837977	&0.383296		&1.050847	\\
53	&1	&5	&93.918767	&0.844116	&0.392687		&1.062329	\\
54	&1	&100	&93.803809	&0.843159	&0.392971		&1.097550	\\
55	&5	&100	&93.693515	&0.843212	&0.393245		&1.104341	\\
56	&1	&10	&93.390166	&0.839254	&0.393999		&1.277776	\\
57	&25	&50	&94.055162	&0.836221	&0.392350		&1.356146	\\
58	&1	&50	&93.086849	&0.835691	&0.394758		&1.442529	\\
59	&1	&25	&93.007858	&0.834618	&0.394956		&1.497993	\\
60	&10	&50	&92.099608	&0.835108	&0.397256		&1.537372	\\
61	&5	&50	&92.015427	&0.835055	&0.397471		&1.546178	\\
62	&5	&10	&91.911263	&0.834438	&0.397738		&1.581408	\\
63	&5	&25	&91.546009	&0.832125	&0.398677		&1.713465	\\
64	&10	&25	&91.092644	&0.831282	&0.399852		&1.786147	\\
65	&0.1&5	&137.750793	&0.898311	&0.314283		&3.917614\\	
66  &0.1&1	&280.002690	&0.969840	&0.211255		&large		
\label{3S1combinations}	
\end{longtable}
\textbf{Discussion:}
\begin{itemize}
\item From the above 66 combinations, 64 of them have $MAE < 2$. The average value for $r_m$ from these 64 combinations is determined to be 0.8427 fm. 
\item Keeping $r_m$ fixed, one needs to vary only one parameter $a_m$ (because $V_0$ is dependent on $a_m$). So, only $^{12}C_1$, \textit{i.e.} only one of the energies from the data set needs to be considered for optimising the parameter $a_m$.
\item  Utilizing the optimised model parameters, the $^3S_1$ wave function has been determined using equation \ref{wfn}.
\item Then, the proprotionality factor for $^3D_1$  wavefunction has been determined such that quadrupole moment $Q_D = 0.2589 fm^2$ is obtained using Eq. \ref{quad} and deuteron wave function (DWF) from Eq. \ref{dwf} is normalised. 
\item Utilizing the obtained DWF, other static properties of deuteron have been determined and tabulated for all 12 energies in Table \ref{3S1final}.
\item One can observe that the depth of the potential keeps increasing with increasing energy except for the first two values. This might be because the first data point is added from Arndt data and is not part of mean energy analysis data of Granada.
\item The values of depth $V_0$ and width $a_m$, given in bold, are utilised for obtaining possible range of parameters in our calculations.
\end{itemize}
\begin{table}
\centering
\caption{Setting $r_m = 0.8427 fm$, value of $a_m$ is optimised for all 12 energy data points. Corresponding $V_0$ for each value of $a_m$ is determined. Retaining experimental value of quadrupole moment, the proportionality constant is determined such that DWF is normalised. The static properties determined from the obtained DWF are calculated.}
\scalebox{0.9}{
\begin{tabular}{c c c c c c c c c} 
\hline
\multirow{2}{*}{\begin{tabular}[c]{@{}c@{}}$E$\\(MeV)\end{tabular}} & \multirow{2}{*}{\begin{tabular}[c]{@{}c@{}}$V_0$\\(MeV)\end{tabular}}   & \multirow{2}{*}{\begin{tabular}[c]{@{}c@{}}$a_m$\\(fm)\end{tabular}} & \multirow{2}{*}{\begin{tabular}[c]{@{}c@{}}$r_{Dm}$\\(fm)\end{tabular}} & \multirow{2}{*}{\begin{tabular}[c]{@{}c@{}}$r_{ch}$\\(fm)\end{tabular}} &\multirow{2}{*}{\begin{tabular}[c]{@{}c@{}}$\mu_D$\\($\mu_N$)\end{tabular}} & \multirow{2}{*}{\begin{tabular}[c]{@{}c@{}}$a_t$\\(fm)\end{tabular}}  & \multirow{2}{*}{\begin{tabular}[c]{@{}c@{}}$r_t$\\(fm)\end{tabular}} & \begin{tabular}[c]{@{}c@{}}\textbf{Overall }\\\textbf{MAE}\end{tabular}\\ 
\\
\hline
0.1& 98.6121  & 0.3816  & 1.9557 & 2.1106   & 0.8685      & 5.3886 & 1.7702 &  0.8080 \\
1& 93.7538  & 0.3931  & 1.9582 & 2.1129   & 0.8686     & 5.4076 & 1.7786 &  1.1180  \\
5& \textbf{93.5772}  & \textbf{0.3935}  & 1.9583 & 2.1130   & 0.8686 & 5.4084 & 1.7790 & 1.1320  \\
10    & 94.6645  & 0.3909  & 1.9577 & 2.1125  & 0.8686        & 5.4035 & 1.7770 & 1.0500   \\
25    & 99.3922  & 0.3798  & 1.9553 & 2.1103  &  0.8685       & 5.3861 & 1.7689 & 0.7630  \\
50    & 113.7154 & 0.3510  & 1.9491 & 2.1045  &  0.8683       & 5.3560 & 1.7493 &  0.3480    \\
100   & 94.7093  & 0.3907  & 1.9577 & 2.1125  &  0.8686      & 5.4034 & 1.7769 & 1.0460 \\
150   & 102.9116 & 0.3721  & 1.9536 & 2.1088  &  0.8685      & 5.3762 & 1.7635 & 0.5780  \\
200   & 107.2748 & 0.3632  & 1.9517 & 2.1070  &  0.8684      & 5.3667 & 1.7574 & 0.4170  \\
250   & 110.7654 & 0.3565  & 1.9503 & 2.1056  &  0.8684       & 5.3605 & 1.7529 &  0.3470  \\
300   & 113.7755 & 0.3509  & 1.9491 & 2.1045  &  0.8683     & 5.3559 & 1.7492 & 0.3480  \\
350   & \textbf{116.3823} & \textbf{0.3464} & 1.9481 & 2.1036  & 0.8683      & 5.3524 & 1.7461 & 0.4140  \\ 
\hline
 & &   &      &      & &    &               \\ 
&& Avg.=  & 1.95371& 2.10882 & 0.86847  & 5.38046& 1.76408 & 0.69738   \\ 
&& St. Dev.=   & 0.00392& 0.00363 & 0.00010  & 0.02188& 0.01263                                                              & 0.32584 \\
\hline
\end{tabular}}
\label{3S1final}
\end{table}



\subsection{Singlet State Analysis:}
\begin{itemize}
\item Since, we have three parameters to be determined for $^1S_0$ state, a total of 220 combinations need to considered. All these are shown in Table \ref{220combi}, where the data have been once again presented in ascending order of overall MAE.
\item Out of these 220, only 130 of them have $MAE < 2$. The average value for $r_m$ from these 130 combinations is determined to be 0.897 fm. 
\item Keeping $r_m$ fixed, one needs to vary only two parameter $V_0$ and $a_m$. So, only $^{12}C_2$, that is 66 combinations need to be worked out. These are given in Table 6.
\item A total of 14 combinations are having $MAE < 1$. These have been considered for determining energy scattering parameters ($a_s$ and $r_s$) and are shown in Table \ref{1S0final}.
\item The values of depth $V_0$ and width $a_m$, given in bold, are utilised for obtaining possible range of values for scattering parameters in our calculations.
\end{itemize}

{\small
\begin{longtable}{cccccccc}
\caption{\textbf{$^1S_0$ state:} Model parameters for 220 combinations, each with three lab energies and obtained by minimising MSE. The overall MAE is determined by obtaining SPS for remaining experimental data points. The data is sorted with respect to MAE in ascending order.}\\
\hline
\textbf{\small{Sr. No.}} & \multirow{2}{*}{\begin{tabular}[c]{@{}c@{}}$E_1$\\(MeV)\end{tabular}}  & \multirow{2}{*}{\begin{tabular}[c]{@{}c@{}}$E_2$\\(MeV)\end{tabular}}  & \multirow{2}{*}{\begin{tabular}[c]{@{}c@{}}$E_3$\\(MeV)\end{tabular}}  & \multirow{2}{*}{\begin{tabular}[c]{@{}c@{}}$V_0$\\(MeV)\end{tabular}}  & \multirow{2}{*}{\begin{tabular}[c]{@{}c@{}}$r_m$\\(fm)\end{tabular}} & \multirow{2}{*}{\begin{tabular}[c]{@{}c@{}}$a_m$\\(fm)\end{tabular}} & \multirow{2}{*}{\begin{tabular}[c]{@{}c@{}}\textbf{Overall}\\\textbf{MAE}\end{tabular}}  \\ 
\\
\hline
\endfirsthead
\multicolumn{8}{c}%
{\tablename\ \thetable\ -- \textit{Continued from previous page}} \\
\hline
\textbf{\small{Sr. No.}} & \multirow{2}{*}{\begin{tabular}[c]{@{}c@{}}$E_1$\\(MeV)\end{tabular}}  & \multirow{2}{*}{\begin{tabular}[c]{@{}c@{}}$E_2$\\(MeV)\end{tabular}}  & \multirow{2}{*}{\begin{tabular}[c]{@{}c@{}}$E_3$\\(MeV)\end{tabular}}  & \multirow{2}{*}{\begin{tabular}[c]{@{}c@{}}$V_0$\\(MeV)\end{tabular}}  & \multirow{2}{*}{\begin{tabular}[c]{@{}c@{}}$r_m$\\(fm)\end{tabular}} & \multirow{2}{*}{\begin{tabular}[c]{@{}c@{}}$a_m$\\(fm)\end{tabular}} & \multirow{2}{*}{\begin{tabular}[c]{@{}c@{}}\textbf{Overall}\\\textbf{MAE}\end{tabular}}  \\
\\ 
\hline
\endhead
\hline \multicolumn{8}{r}{\textit{Continued on next page}} \\
\endfoot
\hline
\endlastfoot
1	&1	&50	&250	&69.106747	&0.901205	&0.375310		&0.693171		\\
2	&1	&50	&300	&71.421730	&0.901418	&0.369307		&0.696128		\\
3	&0.1	&50	&250	&68.412687	&0.903463	&0.377581		&0.717443		\\
4	&1	&100	&250	&79.682938	&0.880680	&0.350055		&0.717933		\\
5	&1	&100	&300	&81.987577	&0.879532	&0.345220		&0.719582		\\
6	&0.1	&50	&300	&70.731083	&0.903718	&0.371473		&0.721345		\\
7	&0.1	&200	&250	&74.094822	&0.890727	&0.363091		&0.744621		\\
8	&1	&100	&350	&83.268723	&0.878941	&0.342619		&0.750327		\\
9	&0.1	&100	&250	&79.064980	&0.882510	&0.351841		&0.751583		\\
10	&0.1	&100	&300	&81.387358	&0.881353	&0.346910		&0.754296		\\
11	&1	&50	&350	&72.979896	&0.901578	&0.365433		&0.765976		\\
12	&1	&100	&200	&76.202160	&0.882633	&0.357774		&0.782170		\\
13	&0.1	&100	&350	&82.694478	&0.880750	&0.344227		&0.786829		\\
14	&0.1	&50	&350	&72.301140	&0.903906	&0.367510		&0.794040		\\
15	&1	&150	&250	&86.814774	&0.869893	&0.335703		&0.805717		\\
16	&1	&50	&200	&65.939275	&0.900949	&0.384056		&0.813704		\\
17	&1	&25	&250	&60.825960	&0.922498	&0.399589		&0.814505		\\
18	&1	&150	&300	&88.776301	&0.868372	&0.332064		&0.815033		\\
19	&0.1	&100	&200	&75.575474	&0.884471	&0.359676		&0.815752		\\
20	&1	&150	&200	&83.391554	&0.872785	&0.342358		&0.821358		\\
21	&1	&150	&350	&89.510470	&0.867826	&0.330733		&0.828324		\\
22	&1	&25	&300	&63.004801	&0.924249	&0.392714		&0.831579		\\
23	&0.1	&25	&250	&60.049341	&0.925329	&0.402465		&0.837142		\\
24	&0.1	&50	&200	&65.250191	&0.903142	&0.386458		&0.839804		\\
25	&0.1	&150	&250	&86.238087	&0.871539	&0.337260		&0.846859		\\
26	&1	&200	&250	&88.940429	&0.866840	&0.331768		&0.854196		\\
27	&0.1	&25	&300	&62.221466	&0.927187	&0.395468		&0.855629		\\
28	&0.1	&150	&300	&88.230758	&0.869983	&0.333526		&0.857504		\\
29	&0.1	&150	&200	&82.771862	&0.874495	&0.344072		&0.861381		\\
30	&0.1	&150	&350	&89.003054	&0.869406	&0.332113		&0.872101		\\
31	&1	&200	&300	&92.791959	&0.862606	&0.324967		&0.902234		\\
32	&1	&200	&350	&92.810359	&0.862588	&0.324936		&0.902573		\\
33	&1	&300	&350	&92.858437	&0.862515	&0.324854		&0.904029		\\
34	&0.1	&250	&300	&91.173683	&0.865407	&0.328232		&0.930979		\\
35	&1	&250	&350	&93.735225	&0.861191	&0.323366		&0.932878		\\
36	&200	&300	&350	&92.977982	&0.862052	&0.324478		&0.943724		\\
37	&1	&100	&150	&71.426516	&0.885796	&0.369286		&0.944335		\\
38	&1	&25	&350	&64.580624	&0.925441	&0.387978		&0.948199		\\
39	&0.1	&200	&300	&92.299268	&0.864088	&0.326278		&0.950222		\\
40	&0.1	&200	&350	&92.365799	&0.864024	&0.326163		&0.951504		\\
41	&1	&250	&300	&94.487069	&0.860334	&0.322107		&0.952344		\\
42	&0.1	&300	&350	&92.795803	&0.863383	&0.325425		&0.965462		\\
43	&1	&25	&200	&57.965254	&0.919940	&0.409270		&0.969674		\\
44	&5	&50	&250	&71.827108	&0.892874	&0.366835		&0.973636		\\
45	&0.1	&25	&350	&63.800001	&0.928452	&0.390630		&0.976690		\\
46	&0.1	&100	&150	&70.797575	&0.887641	&0.371345		&0.979252		\\
47	&5	&50	&300	&74.123940	&0.892922	&0.361222		&0.981621		\\
48	&0.1	&250	&350	&93.354450	&0.862521	&0.324475		&0.984725		\\
49	&0.1	&25	&200	&57.204023	&0.922609	&0.412304		&0.995303		\\
50	&5	&25	&250	&63.895898	&0.912060	&0.388851		&1.004328		\\
51	&5	&25	&300	&66.094065	&0.913414	&0.382434		&1.022167		\\
52	&5	&50	&350	&75.629225	&0.892977	&0.357687		&1.046042		\\
53	&5	&100	&250	&82.103049	&0.873876	&0.343354		&1.057131		\\
54	&5	&100	&300	&84.333908	&0.872755	&0.338877		&1.064852		\\
55	&5	&50	&200	&68.642373	&0.892874	&0.375101		&1.074362		\\
56	&5	&100	&350	&85.507203	&0.872206	&0.336593		&1.094027		\\
57	&5	&100	&200	&78.658031	&0.875816	&0.350645		&1.104883		\\
58	&1	&10	&250	&52.927823	&0.948943	&0.428027		&1.111395		\\
59	&1	&50	&150	&61.827170	&0.900627	&0.396478		&1.119611		\\
60	&5	&25	&350	&67.653891	&0.914328	&0.378085		&1.126204		\\
61	&5	&10	&250	&56.385130	&0.935358	&0.414085		&1.138910		\\
62	&5	&25	&200	&60.977855	&0.910107	&0.397952		&1.139250		\\
63	&0.1	&10	&250	&52.077895	&0.952562	&0.431693		&1.144643		\\
64	&0.1	&50	&150	&61.149255	&0.902712	&0.399064		&1.151779		\\
65	&1	&10	&300	&54.942606	&0.952551	&0.420093		&1.153683		\\
66	&5	&150	&250	&89.075067	&0.863741	&0.329838		&1.166520		\\
67	&5	&150	&200	&85.739120	&0.866489	&0.336071		&1.170322		\\
68	&5	&10	&300	&58.457941	&0.938219	&0.406721		&1.174324		\\
69	&5	&150	&300	&90.908777	&0.862348	&0.326560		&1.179875		\\
70	&0.1	&10	&300	&54.076475	&0.956371	&0.423609		&1.189623		\\
71	&5	&150	&350	&91.486232	&0.861926	&0.325549		&1.191246		\\
72	&5	&100	&150	&73.889807	&0.878977	&0.361587		&1.235965		\\
73	&10	&25	&250	&66.922285	&0.902818	&0.379140		&1.253315		\\
74	&5	&200	&250	&93.523254	&0.858177	&0.322026		&1.255487		\\
75	&5	&300	&350	&94.074841	&0.857910	&0.321101		&1.256449		\\
76	&5	&200	&350	&94.534465	&0.857221	&0.320331		&1.268931		\\
77	&5	&200	&300	&94.713008	&0.857055	&0.320034		&1.273181		\\
78	&10	&25	&300	&69.137374	&0.903791	&0.373121		&1.274380		\\
79	&5	&250	&350	&95.202504	&0.856234	&0.319221		&1.292186		\\
80	&1	&10	&200	&50.350864	&0.943580	&0.439036		&1.293872		\\
81	&5	&10	&200	&53.708317	&0.931144	&0.424339		&1.303345		\\
82	&5	&250	&300	&96.176589	&0.855144	&0.317628		&1.319085		\\
83	&5	&10	&350	&59.999123	&0.940181	&0.401528		&1.327678		\\
84	&0.1	&10	&200	&49.528198	&0.946884	&0.442899		&1.330448		\\
85	&1	&10	&350	&56.466435	&0.955028	&0.414425		&1.330920		\\
86	&5	&50	&150	&64.488669	&0.892965	&0.386853		&1.338231		\\
87	&1	&5	&250	&49.141483	&0.964478	&0.444134		&1.341103		\\
88	&10	&50	&250	&74.523655	&0.885352	&0.359039		&1.348966		\\
89	&10	&50	&300	&76.799822	&0.885227	&0.353775		&1.365209		\\
90	&10	&25	&200	&63.949257	&0.901428	&0.387746		&1.365928		\\
91	&10	&25	&350	&70.671879	&0.904449	&0.369127		&1.368325		\\
92	&0.1	&10	&350	&55.593733	&0.958987	&0.417820		&1.373717		\\
93	&1	&25	&150	&54.328208	&0.916027	&0.422868		&1.386019		\\
94	&0.1	&5	&250	&48.286883	&0.968468	&0.448157		&1.388205		\\
95	&1	&5	&300	&51.068909	&0.969165	&0.435558		&1.399746		\\
96	&0.1	&25	&150	&53.591393	&0.918421	&0.426127		&1.421470		\\
97	&10	&50	&200	&71.319582	&0.885621	&0.366891		&1.425465		\\
98	&10	&50	&350	&78.244939	&0.885176	&0.350556		&1.426545		\\
99	&0.1	&5	&300	&50.193386	&0.973414	&0.439421		&1.450523		\\
100	&5	&25	&150	&57.253135	&0.907190	&0.410716		&1.505230		\\
101	&10	&100	&250	&84.530724	&0.867581	&0.337059		&1.530471		\\
102	&1	&5	&200	&46.704006	&0.957429	&0.455991		&1.535101		\\
103	&10	&100	&300	&86.682173	&0.866471	&0.332917		&1.547869		\\
104	&10	&100	&200	&81.120042	&0.869543	&0.343968		&1.555326		\\
105	&10	&100	&350	&87.737968	&0.865960	&0.330941		&1.577003		\\
106	&0.1	&5	&200	&45.881834	&0.961008	&0.460231		&1.585111		\\
107	&1	&5	&350	&52.550988	&0.972387	&0.429363		&1.612378		\\
108	&10	&50	&150	&67.116829	&0.886133	&0.378088		&1.641743		\\
109	&10	&100	&150	&76.348517	&0.872760	&0.354429		&1.645180		\\
110	&10	&150	&200	&88.114197	&0.860608	&0.330107		&1.668178		\\
111	&0.1	&5	&350	&51.664682	&0.976813	&0.433096		&1.671395		\\
112	&10	&25	&150	&60.136983	&0.899409	&0.399815		&1.680255		\\
113	&10	&150	&250	&91.356167	&0.857975	&0.324270		&1.682730		\\
114	&1	&50	&100	&56.580145	&0.900006	&0.414509		&1.692123		\\
115	&0.1	&1	&250	&44.774617	&0.985427	&0.465320		&1.697181		\\
116	&10	&150	&300	&93.051394	&0.856699	&0.321344		&1.702910		\\
117	&10	&150	&350	&93.457563	&0.856403	&0.320654		&1.712382		\\
118	&0.1	&1	&300	&47.183981	&0.988486	&0.453023		&1.725120		\\
119	&0.1	&50	&100	&55.913525	&0.901886	&0.417403		&1.737669		\\
120	&10	&300	&350	&95.245615	&0.853673	&0.317637		&1.768202		\\
121	&5	&10	&150	&50.337106	&0.924571	&0.438730		&1.772785		\\
122	&10	&200	&250	&95.638373	&0.852761	&0.316978		&1.790223		\\
123	&10	&200	&350	&96.244558	&0.852196	&0.315989		&1.801040		\\
124	&10	&200	&300	&96.635601	&0.851837	&0.315356		&1.811839		\\
125	&10	&250	&350	&96.641823	&0.851618	&0.315341		&1.816855		\\
126	&1	&10	&150	&47.125994	&0.935019	&0.454540		&1.820187		\\
127	&5	&50	&100	&59.203950	&0.893099	&0.403762		&1.831080		\\
128	&10	&250	&300	&97.853486	&0.850276	&0.313399		&1.855489		\\
129	&0.1	&10	&150	&46.342521	&0.937771	&0.458698		&1.869466		\\
130	&0.1	&1	&200	&42.528944	&0.975860	&0.478255		&1.899805		\\
131	&0.1	&1	&350	&48.012793	&0.995775	&0.448966		&2.037688		\\
132	&10	&50	&100	&61.788933	&0.886985	&0.394064		&2.048093		\\
133	&1	&5	&150	&43.680037	&0.945909	&0.472774		&2.116530		\\
134	&25	&50	&250	&80.316066	&0.871277	&0.344007		&2.156406		\\
135	&25	&50	&200	&77.056537	&0.872166	&0.351152		&2.176841		\\
136	&0.1	&5	&150	&42.905221	&0.948749	&0.477340		&2.179469		\\
137	&25	&50	&300	&82.532646	&0.870768	&0.339404		&2.193151		\\
138	&1	&25	&100	&49.702833	&0.909077	&0.442912		&2.202878		\\
139	&5	&25	&100	&52.517687	&0.902198	&0.429327		&2.218603		\\
140	&25	&50	&350	&83.815921	&0.870503	&0.336827		&2.249532		\\
141	&0.1	&25	&100	&48.998055	&0.910918	&0.446557		&2.260146		\\
142	&25	&50	&150	&72.702062	&0.873641	&0.361489		&2.286375		\\
143	&10	&25	&100	&55.297933	&0.896086	&0.417240		&2.290449		\\
144	&25	&50	&100	&67.209480	&0.876017	&0.376072		&2.506611		\\
145	&25	&100	&200	&86.551281	&0.857386	&0.330660		&2.573182		\\
146	&200	&250	&350	&98.501016	&0.845976	&0.310573		&2.598465		\\
147	&25	&100	&250	&89.888162	&0.855294	&0.324463		&2.604920		\\
148	&25	&100	&300	&91.833871	&0.854185	&0.321015		&2.645128		\\
149	&25	&100	&350	&92.580844	&0.853779	&0.319722		&2.672145		\\
150	&0.1	&1	&150	&38.535909	&0.964784	&0.505175		&2.704045		\\
151	&5	&10	&100	&46.071588	&0.912570	&0.460216		&2.722212		\\
152	&25	&150	&200	&93.424904	&0.848956	&0.317999		&2.820046		\\
153	&150	&300	&350	&97.422176	&0.846271	&0.311553		&2.835157		\\
154	&25	&150	&250	&96.438863	&0.846505	&0.312962		&2.880206		\\
155	&1	&10	&100	&43.074535	&0.918771	&0.477999		&2.894830		\\
156	&25	&300	&350	&97.580422	&0.845751	&0.311126		&2.905650		\\
157	&25	&150	&350	&97.739224	&0.845512	&0.310864		&2.912900		\\
158	&25	&150	&300	&97.776269	&0.845484	&0.310805		&2.914294		\\
159	&0.1	&10	&100	&42.351901	&0.920312	&0.482671		&2.972663		\\
160	&25	&250	&350	&99.625324	&0.842731	&0.307816		&3.011398		\\
161	&25	&200	&350	&99.898925	&0.842338	&0.307381		&3.026484		\\
162	&25	&200	&250	&100.329734	&0.841937	&0.306714		&3.042687		\\
163	&25	&200	&300	&100.833177	&0.841474	&0.305940		&3.062787		\\
164	&25	&250	&300	&101.437726	&0.840715	&0.305006		&3.093527		\\
165	&1	&5	&100	&39.939787	&0.923124	&0.498547		&3.321657		\\
166	&0.1	&5	&100	&39.241889	&0.924273	&0.503697		&3.413964		\\
167	&10	&25	&50	&48.427383	&0.887659	&0.447909		&3.492464		\\
168	&50	&100	&200	&93.036366	&0.845318	&0.316854		&3.587187		\\
169	&5	&25	&50	&45.773686	&0.888988	&0.462923		&3.652452		\\
170	&50	&100	&250	&96.235982	&0.843060	&0.311435		&3.677018		\\
171	&50	&100	&300	&97.831504	&0.842018	&0.308842		&3.732774		\\
172	&50	&100	&350	&98.081528	&0.841858	&0.308441		&3.744466		\\
173	&50	&300	&350	&99.619316	&0.839351	&0.305829		&3.870852		\\
174	&1	&25	&50	&43.137635	&0.889643	&0.480020		&3.871586		\\
175	&0.1	&1	&100	&36.464046	&0.926864	&0.525711		&3.895002		\\
176	&200	&250	&300	&104.136588	&0.834181	&0.299196		&3.966924		\\
177	&0.1	&25	&50	&42.493290	&0.889644	&0.484605		&3.978933		\\
178	&50	&150	&200	&99.845025	&0.837125	&0.305228		&3.995551		\\
179	&100	&300	&350	&100.093987	&0.837923	&0.304643		&4.093311		\\
180	&50	&150	&250	&102.475366	&0.834914	&0.301162		&4.096088		\\
181	&50	&150	&350	&102.506158	&0.834888	&0.301115		&4.097374		\\
182	&50	&250	&350	&102.547580	&0.834826	&0.301049		&4.100770		\\
183	&150	&250	&350	&102.593867	&0.834705	&0.300945		&4.118219		\\
184	&50	&150	&300	&103.231471	&0.834307	&0.300024		&4.135096		\\
185	&50	&200	&350	&103.766970	&0.833032	&0.299129		&4.201686		\\
186	&50	&250	&300	&105.211529	&0.831721	&0.296992		&4.277158		\\
187	&50	&200	&300	&105.512613	&0.831338	&0.296539		&4.298649		\\
188	&50	&200	&250	&105.772235	&0.831093	&0.296160		&4.313871		\\
189	&5	&10	&50	&40.165668	&0.877385	&0.501389		&4.722662		\\
190	&100	&250	&350	&104.408473	&0.830149	&0.297011		&4.758500		\\
191	&150	&250	&300	&107.704509	&0.826297	&0.292095		&4.992494		\\
192	&150	&200	&350	&106.742895	&0.826545	&0.293295		&5.014595		\\
193	&0.1	&1	&10	&37.470535	&0.872896	&0.525527		&5.078188		\\
194	&100	&200	&350	&107.299645	&0.825389	&0.292248		&5.155127		\\
195	&1	&10	&50	&37.696821	&0.867371	&0.524343		&5.182563		\\
196	&100	&250	&300	&108.520852	&0.824605	&0.290556		&5.209737		\\
197	&100	&150	&350	&107.819985	&0.824568	&0.291418		&5.223816		\\
198	&100	&150	&200	&109.023715	&0.823444	&0.289687		&5.307572		\\
199	&0.1	&10	&50	&37.136204	&0.864047	&0.530371		&5.324111		\\
200	&100	&150	&300	&109.962564	&0.822590	&0.288360		&5.372220		\\
201	&100	&150	&250	&110.559310	&0.822058	&0.287527		&5.415458		\\
202	&100	&200	&300	&110.551876	&0.821786	&0.287478		&5.437664		\\
203	&150	&200	&300	&111.206513	&0.820635	&0.286372		&5.572116		\\
204	&100	&200	&250	&112.391358	&0.819871	&0.284893		&5.592481		\\
205	&1	&5	&50	&35.452204	&0.844164	&0.551371		&5.910267		\\
206	&150	&200	&250	&114.504908	&0.816665	&0.281599		&5.945619		\\
207	&0.1	&5	&50	&35.000441	&0.837637	&0.558131		&6.065017		\\
208	&0.1	&1	&50	&33.547357	&0.797536	&0.587618		&6.790136		\\
209	&5	&10	&25	&36.602700	&0.797587	&0.551225		&6.903455		\\
210	&250	&300	&350	&87.833860	&0.884140	&0.342317		&7.298117		\\
211	&1	&10	&25	&35.808537	&0.734105	&0.583234		&7.668749		\\
212	&0.1	&10	&25	&35.837910	&0.712972	&0.591659		&7.869697		\\
213	&1	&5	&25	&37.735829	&0.592699	&0.623013		&8.691153		\\
214	&0.1	&5	&25	&38.819828	&0.546812	&0.632938		&8.893464		\\
215	&0.1	&5	&10	&41.048833	&0.486927	&0.638220		&9.223375		\\
216	&0.1	&1	&5	&47.901617	&0.309455	&0.662032		&9.684701		\\
217	&0.1	&1	&25	&52.566843	&0.200082	&0.678402		&9.772956		\\
218	&1	&5	&10	&59.902374	&0.108433	&0.672882		&10.157729		\\
219	&25	&100	&150	&14.631010	&2.000000	&1.721783		&27.052529		\\
220	&50	&100	&150	&18.787788	&2.000000	&1.524581		&27.920139		
\label{220combi}
\end{longtable}
}

{\small
\begin{longtable}{cccccc}
\caption{Model parameters for 66 combinations for $^{1}S_{0}$ state. After fixing $r_m$=0.897fm from 130 combinations, two parameters $V_0$ and $a_m$ produces $^{12}C_2$ \textit{i.e.} 66 combinations. The data has been sorted with ascending values of MAE.}\\
\hline
\textbf{\small{Sr. No.}} & \multirow{2}{*}{\begin{tabular}[c]{@{}c@{}}$E_1$\\(MeV)\end{tabular}} & \multirow{2}{*}{\begin{tabular}[c]{@{}c@{}}$E_2$\\(MeV)\end{tabular}} & \multirow{2}{*}{\begin{tabular}[c]{@{}c@{}}$V_0$\\(MeV)\end{tabular}}  & \multirow{2}{*}{\begin{tabular}[c]{@{}c@{}}$a_m$\\(fm)\end{tabular}} & \multirow{2}{*}{\begin{tabular}[c]{@{}c@{}}\textbf{Overall}\\\textbf{MAE}\end{tabular}}  \\ 
\\
\hline
\endfirsthead
\multicolumn{6}{c}%
{\tablename\ \thetable\ -- \textit{Continued from previous page}} \\
\hline
\textbf{\small{Sr. No.}} & \multirow{2}{*}{\begin{tabular}[c]{@{}c@{}}$E_1$\\(MeV)\end{tabular}} & \multirow{2}{*}{\begin{tabular}[c]{@{}c@{}}$E_2$\\(MeV)\end{tabular}} & \multirow{2}{*}{\begin{tabular}[c]{@{}c@{}}$V_0$\\(MeV)\end{tabular}}  & \multirow{2}{*}{\begin{tabular}[c]{@{}c@{}}$a_m$\\(fm)\end{tabular}} & \multirow{2}{*}{\begin{tabular}[c]{@{}c@{}}\textbf{Overall}\\\textbf{MAE}\end{tabular}}  \\ 
\\
\hline
\endhead
\hline \multicolumn{6}{r}{\textit{Continued on next page}} \\
\endfoot
\hline
\endlastfoot
1	&1	&250	&70.872343	&0.370701	&0.698467\\
2	&1	&300	&73.177654	&0.364944	&0.701078\\
3	&0.1	&250	&71.206545	&0.370272	&0.728250   \\
4	&0.1	&300	&73.471352	&0.364650	&0.731490   \\
5	&1	&350	&74.710829	&0.361268	&0.763778   \\
6	&0.1	&350	&74.975596	&0.361061	&0.792975   \\
7	&50	&250	&70.341355	&0.371382	&0.829633   \\
8	&0.1	&200	&68.127003	&0.378378	&0.833878   \\
9	&50	&300	&72.692837	&0.365428	&0.839807   \\
10	&1	&200	&68.136245	&0.377212	&0.909084   \\
11	&50	&350	&74.261240	&0.361617	&0.910795   \\
12	&50	&200	&67.119046	&0.380059	&0.936133   \\
13	&5	&250	&69.807418	&0.372069	&0.982129   \\
14	&5	&300	&72.232346	&0.365888	&0.991642   \\
15	&5	&350	&73.853219	&0.361933	&1.063794   \\
16	&1	&150	&63.714747	&0.390628	&1.083160   \\
17	&5	&200	&66.472351	&0.381146	&1.092938   \\
18	&50	&150	&62.939878	&0.392355	&1.218870   \\
19	&10	&250	&69.126948	&0.372946	&1.285511   \\
20	&10	&300	&71.615379	&0.366503	&1.308314   \\
21	&10	&200	&65.698368	&0.382454	&1.383487   \\
22	&10	&350	&73.286054	&0.362371	&1.392455   \\
23	&25	&250	&68.880906	&0.373263	&1.424039   \\
24	&25	&300	&71.365270	&0.366752	&1.469937   \\
25	&25	&200	&65.472662	&0.382838	&1.490554   \\
26	&25	&350	&73.040584	&0.362560	&1.570751   \\
27	&10	&150	&61.127074	&0.396483	&1.676120   \\
28	&25	&150	&60.966260	&0.396855	&1.736364   \\
29	&50	&100	&57.564358	&0.410362	&1.750752   \\
30	&10	&25	&56.902371	&0.411159	&2.078685   \\
31	&100	&150	&67.728694	&0.381998	&2.234713   \\
32	&25	&100	&54.710528	&0.419695	&2.277759   \\
33	&100	&200	&71.558814	&0.372763	&2.613412   \\
34	&10	&50	&50.448065	&0.438237	&2.769191   \\
35	&1	&50	&49.199901	&0.445781	&2.770181   \\
36	&5	&25	&49.341756	&0.443832	&2.848806   \\
37	&100	&250	&74.510279	&0.366064	&2.941749   \\
38	&0.1	&50	&47.133223	&0.456632	&3.087928   \\
39	&5	&50	&107.547213	&0.301282	&3.098975   \\
40	&100	&300	&76.600568	&0.361517	&3.214328   \\
41	&1	&25	&45.056329	&0.467687	&3.333094   \\
42	&100	&350	&77.904150	&0.358759	&3.418888   \\
43	&0.1	&25	&44.190202	&0.473184	&3.479983   \\
44	&5	&10	&42.746017	&0.480957	&3.760655   \\
45	&1	&10	&39.998919	&0.500915	&4.073499   \\
46	&0.1	&10	&39.374437	&0.505921	&4.171207   \\
47	&1	&5	&37.598535	&0.520116	&4.449863   \\
48	&0.1	&5	&37.042219	&0.525108	&4.533947   \\
49	&0.1	&1	&34.937980	&0.544839	&4.889577   \\
50	&150	&200	&75.760578	&0.366102	&5.164170   \\
51	&0.1	&100	&26.802146	&0.652930	&6.195900   \\
52	&150	&250	&78.477135	&0.361078	&6.197020   \\
53	&1	&100	&26.587903	&0.658229	&6.226937   \\
54	&25	&50	&260.964563	&0.193853	&6.814614   \\
55	&150	&300	&80.288028	&0.357819	&6.901316   \\
56	&5	&100	&25.354859	&0.692074	&7.194710   \\
57	&150	&350	&81.278564	&0.356065	&7.301086   \\
58	&0.1	&150	&14.174935	&1.059441	&8.555006   \\
59	&200	&250	&81.596913	&0.357211	&8.979169   \\
60	&200	&300	&83.072811	&0.355027	&9.897495   \\
61	&200	&350	&83.695050	&0.354113	&10.278621   \\
62	&10	&100	&23.251541	&0.766074	&10.552366   \\
63	&250	&300	&84.777989	&0.353319	&11.606734   \\
64	&250	&350	&84.991275	&0.353061	&11.769382   \\
65	&300	&350	&85.235921	&0.352861	&12.038305   \\
66	&5	&150	&12.499162	&1.305760	&13.119575   \\
\label{66comb}
\end{longtable}}


{\small
\begin{longtable}{cccccccc}
\caption{Setting $r_m = 0.897 fm$, value of $V_0$ and $a_m$ is optimised for all 14 energy data points. Low energy scattering parameters $a_s$ and $r_s$ are further determined for $^1S_0$ state.}\\
\hline
\textbf{\small{Sr. No.}} & \multirow{2}{*}{\begin{tabular}[c]{@{}c@{}}$E_1$\\(MeV)\end{tabular}} & \multirow{2}{*}{\begin{tabular}[c]{@{}c@{}}$E_2$\\(MeV)\end{tabular}} & \multirow{2}{*}{\begin{tabular}[c]{@{}c@{}}$V_0$\\(MeV)\end{tabular}} & \multirow{2}{*}{\begin{tabular}[c]{@{}c@{}}$a_m$\\(fm)\end{tabular}}   & \multirow{2}{*}{\begin{tabular}[c]{@{}c@{}}$a_s$\\(fm)\end{tabular}}& \multirow{2}{*}{\begin{tabular}[c]{@{}c@{}}$r_s$\\(fm)\end{tabular}}& \multirow{2}{*}{\begin{tabular}[c]{@{}c@{}}\textbf{Overall}\\\textbf{MAE}\end{tabular}}  \\ 
\\
\hline
\endfirsthead
\multicolumn{8}{c}%
{\tablename\ \thetable\ -- \textit{Continued from previous page}} \\
\hline
\textbf{\small{Sr. No.}} & \multirow{2}{*}{\begin{tabular}[c]{@{}c@{}}$E_1$\\(MeV)\end{tabular}} & \multirow{2}{*}{\begin{tabular}[c]{@{}c@{}}$E_2$\\(MeV)\end{tabular}} & \multirow{2}{*}{\begin{tabular}[c]{@{}c@{}}$V_0$\\(MeV)\end{tabular}} & \multirow{2}{*}{\begin{tabular}[c]{@{}c@{}}$a_m$\\(fm)\end{tabular}}   & \multirow{2}{*}{\begin{tabular}[c]{@{}c@{}}$a_s$\\(fm)\end{tabular}}& \multirow{2}{*}{\begin{tabular}[c]{@{}c@{}}$r_s$\\(fm)\end{tabular}}& \multirow{2}{*}{\begin{tabular}[c]{@{}c@{}}\textbf{Overall}\\\textbf{MAE}\end{tabular}}  \\ 
\\
\hline
\endhead
\hline \multicolumn{8}{r}{\textit{Continued on next page}} \\
\endfoot
\hline
\endlastfoot
1&0.1&200&68.1270&0.3784&-24.0877&2.4169&0.8339\\
2&0.1&250&71.2065&0.3703&-24.0897&2.4076&0.7282\\
3&0.1&300&73.4714&0.3646&-24.0910&2.4013&0.7315\\
4&0.1&350&\textbf{74.9756}&\textbf{0.3611}&-24.0917&2.3972&0.7930\\
5&1&200&68.1362&0.3772&-21.7968&2.4268&0.9091\\
6&1&250&70.8723&0.3707&-23.1515&2.4124&0.6985\\
7&1&300&73.1777&0.3649&-23.1323&2.4060&0.7011\\
8&1&350&74.7108&0.3613&-23.1200&2.4019&0.7638\\
9&5&250&69.8074&0.3721&-20.5479&2.4280&0.9821\\
10&5&300&72.2323&0.3659&-20.4629&2.4214&0.9916\\
11&50&200&\textbf{67.1191}&\textbf{0.3801}&-21.9063&2.4295&0.9361\\
12&50&250&70.3414&0.3714&-21.7861&2.4201&0.8296\\
13&50&300&72.6928&0.3654&-21.6911&2.4138&0.8398\\
14&50&350&74.2612&0.3616&-21.6234&2.4099&0.9108\\ \hline

	&	&	&	&		Avg.	&-22.5413	&2.4138 & 0.8321\\
	 &   & 	&	&		St. Dev.	& 1.3020	    &0.0104&0.1012
\label{1S0final}
\end{longtable}
}

\begin{table}[H]
\centering
\caption{Obtained $^3S_1$ and $^1S_0$ SPS in comparison with MEPWAD data from \cite{44}. E=0.1 MeV data was taken from Arndt et al. (Private communication).}
\scalebox{1.1}{
\begin{tabular}{ccccc}
\hline 
\multirow{2}{*}{\begin{tabular}[c]{@{}c@{}}E\\(MeV)\end{tabular}} & \multirow{2}{*}{\begin{tabular}[c]{@{}c@{}}$^3S_1$\\\cite{33}\end{tabular}} & \multirow{2}{*}{\begin{tabular}[c]{@{}c@{}}$^3S_1$\\Our\end{tabular}} & \multirow{2}{*}{\begin{tabular}[c]{@{}c@{}}$^1S_0$\\\cite{33}\end{tabular}} & \multirow{2}{*}{\begin{tabular}[c]{@{}c@{}}$^1S_0$\\Our\end{tabular}} 
\\
\\ \hline
0.1$^*$ & 169.320$\pm$0.000 & 169.330$\pm$ 0.041 & 38.430$\pm$0.000 & 36.025$\pm$1.961 \\ 
1 & 147.748$\pm$0.093 & 147.890$\pm$0.123 & 62.105$\pm$0.039 & 61.107$\pm$1.345 \\ 
5 & 118.169$\pm$0.213 & 118.405$\pm$0.210 & 63.689$\pm$0.079 & 64.305$\pm$0.686 \\ 
10 & 102.587$\pm$0.300 & 102.812$\pm$0.235 & 60.038$\pm$0.114 & 61.028$\pm$0.549 \\ 
25 & 80.559$\pm$0.447 & 80.631$\pm$0.220 & 51.011$\pm$0.189 & 51.838$\pm$0.380 \\ 
50 & 62.645$\pm$0.538 & 62.519$\pm$0.118 & 40.644$\pm$0.324 & 40.810$\pm$0.214 \\ 
100 & 43.088$\pm$0.512 & 42.893$\pm$0.207 & 26.772$\pm$0.620 & 26.148$\pm$0.239 \\ 
150 & 30.644$\pm$0.428 & 30.665$\pm$0.531 & 16.791$\pm$0.770 & 16.007$\pm$0.341 \\ 
200 & 21.244$\pm$0.392 & 21.679$\pm$0.865 & 8.759$\pm$0.736 & 8.204$\pm$0.428 \\ 
250 & 13.551$\pm$0.474 & 14.549$\pm$1.203 & 1.982$\pm$0.561 & 1.854$\pm$0.536 \\ 
300 & 6.966$\pm$0.695 & 8.632$\pm$1.543 & -3.855$\pm$0.357 & -3.496$\pm$0.672 \\ 
350 & 1.176$\pm$1.017 & 3.573$\pm$1.885 & -8.923$\pm$0.533 & -8.112$\pm$0.819 \\ 
\hline
\end{tabular}}
\label{table2}
\end{table}
\subsection{Scattering Phase Shifts Data:}
Finally, from the obtained 12 combinations for $^3S_1$ and 14 for $^1S_0$, SPS were determined for each set of optimised model parameters using PFM. Average values of SPS and corresponding uncertainties have been obtained. These are presented in comparision with Granada MEPWAD \cite{33} in Table \ref{table2}.

\end{document}